\documentclass[lettersize,journal]{IEEEtran}
\usepackage{amsmath,amsfonts}
\usepackage{algorithmic}
\usepackage{array}
\usepackage[caption=false,font=normalsize,labelfont=sf,textfont=sf]{subfig}
\usepackage{textcomp}
\usepackage{stfloats}
\usepackage{url}
\usepackage{verbatim}
\usepackage{graphicx}
\usepackage{balance}
\usepackage{amsmath}
\usepackage{amssymb}
\usepackage{amsthm}
\usepackage{mathtools}
\usepackage{enumitem}
\usepackage{color}
\usepackage{cite}

\newtheorem{proposition}{\hspace{0pt}\bf Proposition}

\newtheorem{theorem}{\hspace{0pt}\bf Theorem}
\newtheorem{corollary}{\hspace{0pt}\bf Corollary}

\newtheorem{definition}{\hspace{0pt}\bf Definition}
\newcommand{\argmin}{\operatornamewithlimits{argmin}}

\begin{document}

\title{Stability of Flow Models for Graph Signals}

\author{\IEEEauthorblockN{Martin Schmidt and Gonzalo Mateos}
\thanks{This work was supported in part by NSF under Grant ECCS 2231036. \emph{(Corresponding author: Gonzalo Mateos.)}

Martin Schmidt and Gonzalo Mateos are with the Dept. of Electrical and Computer Engineering, University of Rochester, Rochester, NY 14627, USA (e-mails: mschmi21@ur.rochester.edu; gmateosb@ece.rochester.edu).}}


\maketitle

%
%

\begin{abstract}
Generating signals on graphs requires permutation-equivariant models that exhibit stability with respect to relative structural perturbations. While favorable stability properties of Graph Neural Networks (GNNs) have been well documented, it is unclear how structural errors propagate through the dynamics of continuous generative flow models that are gaining traction for graph signal generation. In this paper, we analyze continuous normalized flow models parameterized by GNNs and show that permutation equivariance is preserved for both the resulting continuous-time ordinary differential equations and their discrete numerical approximations used as graph signal samplers. Our primary contribution is to derive explicit stability bounds on the generated probability distributions, which quantify how relative graph perturbations affect the final sampled signals. Motivated by these theoretical bounds, we introduce a stability-promoting regularized flow matching strategy that actively penalizes the spatial Lipschitz constant of the vector field during model training. Experiments using synthetic smooth signals on stochastic block model graphs and real-world fMRI signals on brain connectomes demonstrate that this bound-oriented approach yields generative models that are more robust to structural noise, without sacrificing output quality.
\end{abstract}


%
%

\section{Introduction}

\IEEEPARstart{G}{\lowercase{enerative}} models are powerful tools for learning underlying data distributions from finite samples. Recent advances in continuous-time generative modeling, including flow matching \cite{lipman2022flow}, diffusion models \cite{ho2020denoisingdiffusionprobabilisticmodels}, and neural ordinary differential equations (ODEs) \cite{chen2019neuralordinarydifferentialequations}, have achieved remarkable success generating realistic images, audio and video. However,  their application to non-Euclidean domains is still developing. Specifically, while considerable effort has gone into generating graph topologies \cite{liu2019graphnormalizingflows, niu2020permutationinvariantgraphgeneration, jo2022scorebasedgenerativemodelinggraphs, liu2023generativediffusionmodelsgraphs, vignac2023digress, chen_2026}, significantly less attention has been paid to generating signals supported on the nodes of a given graph structure \cite{rozada2025graphawarediffusionsignalgeneration,uslu2025graphsignalgenerativediffusion, zhu2024graphsignaldiffusionmodel, yang2025topological}, despite their ubiquity in real-world applications \cite{ortega2018graph}. Such signals arise in diverse contexts, including neural activity over structural brain connectomes \cite{huang2018graph}, traffic flows on transportation networks \cite{yu2017spatio}, sensor measurements in distributed systems \cite{jablonski2017graph}, and power injections in electrical grids \cite{ramakrishna2021grid}. Learning to faithfully generate such signals can then be useful for various reasons, e.g., to produce synthetic samples in low-data regimes, to serve as implicit priors in inverse or reconstruction problems, and to enable simulation of realistic system behavior. For instance, recent works have leveraged graph signal diffusion models for probabilistic forecasting of stock prices to capture uncertainties and tail events \cite{uslu2025graphsignalgenerativediffusion}, modeling correlated user-item interactions for collaborative filtering in recommender systems \cite{zhu2024graphsignaldiffusionmodel}, and optimizing wireless resource allocation \cite{uslu2026graph}. While said application-driven impetus has fueled exciting architectural advances \cite{rozada2025graphawarediffusionsignalgeneration,uslu2025graphsignalgenerativediffusion, zhu2024graphsignaldiffusionmodel, yang2025topological}, the present paper studies fundamental equivariance and stability properties of flow models for graph signal generation. In addition to addressing theoretical questions left unanswered by prior work, the upshot of our analysis has practical implications to model training.\vspace{2pt} 

\noindent\textbf{Permutation equivariance and stability.} Because high-quality data can be both limited and expensive, neural networks must leverage structural priors to learn effectively. To this end, one can e.g., incorporate prior knowledge through data augmentation \cite{alomar2023data} or penalties in the loss function \cite{raissi2019physics}. When possible, arguably the preferred approach is to infuse inductive biases directly into the network architecture \cite{bronstein2021geometric}. \emph{Permutation equivariance} is a non-negotiable desideratum for graph-structured data, 
which guarantees that the generative dynamics natively respect the graph's structural symmetries.

Beyond achieving good performance on training data, machine learning models are expected to generalize well to unseen samples. A key principle underlying such behavior is \emph{stability}: models that respond smoothly to small perturbations in their input tend to generalize better and exhibit improved robustness and transferability \cite{szegedy2013intriguing, bousquet2002stability}; learning with graphs is no exception~\cite{zugner2018adversarial}. Graph-structured data are inherently irregular and often noisy, and in many applications the observed topology is only an imperfect reflection of the underlying system \cite[Ch. 7]{kolaczyk2009book}. For instance, graphs constructed from empirical measurements—such as correlation networks \cite{zalesky2012use} derived from finite-length time series—can vary significantly with the sampling window, leading to noisy graph representations \cite{SI_SPMAG, dong_2019_learning}. Crucially, measuring a model's robustness to such topological noise requires comparing predictive outputs across different graph structures. Because node indexing is often arbitrary, any formal measure of structural stability must be defined modulo permutation \cite{Gama_2020}. Consequently, permutation equivariance is not merely a useful inductive bias, but a strict mathematical prerequisite for analyzing the structural stability of these graph-parametrized learning models.\vspace{2pt}

\noindent\textbf{Innovations in context.} While permutation equivariance, structural stability, and transferability have been rigorously established for Graph Neural Networks (GNNs) \cite{Gama_2020, ruiz2023transferabilitypropertiesgraphneural,kenlay2021interpretablestabilityboundsspectral, levie2021transferability}, these properties have primarily been studied in the context of static, discriminative tasks. In contrast, continuous-time generative models evolve a state trajectory that maps a simple prior distribution to the data distribution. Although related equivariance results exist for flow models \cite{kohler2020equivariant}, in our graph signal processing (GSP) setting equivariance must account for coupled permutations of both the graph signal and the underlying topology. Recent works in graph signal generation have primarily focused on model design rather than theoretical guarantees. In particular, \cite{rozada2025graphawarediffusionsignalgeneration} incorporates the graph structure into a diffusion-style forward process via a heat equation, while \cite{uslu2025graphsignalgenerativediffusion} proposes a U-Net–inspired GNN architecture. However, these approaches do not provide a formal analysis of how structural perturbations propagate through the induced dynamical system. In continuous-time graph signal flows, the evolution is initialized from a random prior and driven by a time-dependent vector field parameterized by a GNN, so that graph-dependent operations are repeatedly applied along the entire trajectory. As a result, even small errors in the underlying graph can accumulate over time and affect the generated samples. Consequently, extending the stability properties of GNNs to this generative setting requires characterizing error propagation through the flow's dynamics. Beyond pointwise stability of the vector field, we are interested in stability of the induced distributions, quantified in terms of Wasserstein distance under perturbations of the graph topology.

In this context, our main contribution is to derive a Wasserstein stability bound that quantifies how relative graph perturbations affect the learnt distribution over graph signals. In support of this main result, we also: (i) formally establish that a continuous normalized flow (CNF) parametrized by a GNN vector field inherits the permutation equivariance of the base architecture; and (ii) propose a regularized flow matching strategy informed by our theoretical findings, which encourages robustness to graph perturbations. Crucially, our bounds demonstrate that the stability of the vector field is intrinsically tied to the graph spectrum, enabling a fundamentally graph-aware regularization strategy.\vspace{2pt}

\noindent\textbf{Paper outline.} The remainder of the paper is organized as follows. Section \ref{sec: preliminaries} reviews the necessary preliminaries. Section \ref{sec: graph flow models} formulates the GNN-parametrized CNF, formally proves their permutation equivariance, and derives the continuous-time stability bounds. Because these continuous models must be evaluated via numerical integration, Section \ref{sec: discrete samplers} extends our theoretical framework to discrete samplers, deriving explicit stability bounds for the Euler and Heun methods. Section \ref{sec: regularization} details the practical implications of these theoretical results, explaining how to leverage our bounds for judicious regularization of the training loss. Section \ref{sec: experiments} numerically illustrates the stability of these models on a synthetic stochastic block model (SBM) graph with smooth signals and a test case of generating fMRI signals supported on a functional brain connectome. Finally, Section \ref{sec: conclusion} concludes the paper by summarizing our findings and discussing limitations of our analysis, with an outlook towards future work. All technical details and proofs are deferred to the appendices.

%
%

\section{Preliminaries}
\label{sec: preliminaries}
This section reviews the preliminary background required to formalize the proposed generative model. We outline the key GSP concepts and survey GNN stability results, followed by the fundamentals of CNFs and flow matching. Throughout the paper, $\| \cdot \|$ denotes the Euclidean norm for vectors, while for matrices, $\| \cdot \|_2$ and $\| \cdot \|_F$ represent the spectral and Frobenius norms, respectively.

%
%

\subsection{Graph Signal Processing}
Let $\mathcal{G}= (\mathcal{V}, \mathcal{E}, \mathcal{W})$ be a known undirected graph with node set $\mathcal{V}$ of cardinality $N$, edge set $\mathcal{E}\subseteq \mathcal{V}\times \mathcal{V}$, and edge weight map $\mathcal{W}: \mathcal{E}\to \mathbb{R}$. Associated with $\mathcal{G}$ is a matrix representation $\mathbf{S}\in \mathbb{R}^{N \times N}$, called a graph shift operator (GSO), which respects the sparsity structure of the graph; specifically, $S_{ij}=0$ if $(i,j)\notin \mathcal{E}$ for $i\neq j$. Common choices for the GSO include the adjacency matrix $\mathbf{A}$, the graph Laplacian $\mathbf{L}$, and their normalized variants. Because $\mathcal{G}$ is undirected then $\mathbf{S}\in\mathcal{S}$ is a symmetric matrix, where $\mathcal{S}=\{\mathbf{S}\in\mathbb{R}^{N\times N}: \mathbf{S}=\mathbf{S}^\top\}$. Denote the GSO eigendecomposition by $\mathbf{S}= \mathbf{V}\mathbf{\Lambda}\mathbf{V}^\top$, where $\mathbf{V} \in \mathbb{R}^{N \times N}$ contains the orthonormal eigenvectors of $\mathbf{S}$ and $\mathbf{\Lambda}$ is the diagonal matrix of corresponding eigenvalues $\{\lambda_i\}_{i=1}^N$. 

Beyond the graph structure itself, we also consider signals defined over the nodes in $\mathcal{V}$.
A graph signal is a function $x: \mathcal{V} \to \mathbb{R}$, mapping each node to a real value.
This function can be represented as a vector $\mathbf{x} \in \mathbb{R}^N$, where $x_i$ denotes the signal value at node $i$. The matrix $\mathbf{V}^\top$ defines the graph Fourier transform (GFT), so that $\bar{\mathbf{x}} = \mathbf{V}^\top \mathbf{x}$ is the GFT of graph signal $\mathbf{x}$~\cite{ortega2018graph}. The eigenvectors $\mathbf{V}$ thus serve as the Fourier basis, while the eigenvalues in $\mathbf{\Lambda}$ are the graph frequencies. 

Graph convolutional filters are defined as polynomial functions of $\mathbf{S}$ parameterized by coefficients $ \{\theta_p\}_{p=0}^{P-1}$. We define a graph filter operator of order $P$ as $\mathbf{H}(\mathbf{S})  \coloneq \sum_{p=0}^{P-1} \theta_p \mathbf{S}^p$, which acts on an input graph signal $\mathbf{x}$ as $\mathbf{z} = \mathbf{H}(\mathbf{S}) \mathbf{x}$ \cite{isufi2024tsp}. Leveraging the GFT definition and the diagonal frequency response of the filter $\mathbf{H}(\mathbf{\Lambda}) = \sum_{p=0}^{P-1} \theta_p \mathbf{\Lambda}^p$, the graph convolution can be written in the spectral domain as $\bar{\mathbf{z}} = \mathbf{H}(\mathbf{\Lambda})\bar{\mathbf{x}}$. Equivalently, the $i$-th spectral component satisfies $\bar{z}_i = h(\lambda_i)\,\bar{x}_i,$
where $h(\lambda):=\sum_{p=0}^{P-1}\theta_p \lambda^p$ is the scalar frequency response of the graph and $\mathbf{H}(\mathbf{\Lambda})=\textrm{diag}(h(\lambda_1),\ldots,h(\lambda_N))$.

A GNN \cite{Gama_2019, kipf2016semi, defferrard2016convolutional} can be constructed from these filters via a cascade of $L$ layers, where each layer entails a graph convolution followed by a pointwise nonlinearity $\sigma(\cdot)$ such as ReLU. At layer $\ell$, let $\mathbf{X}_\ell \in \mathbb{R}^{N \times F_\ell}$ denote the matrix whose columns are the graph signal features, with $\mathbf{X}_0$ and $\mathbf{X}_L$ representing the input and output signals, respectively. For each filter tap $p$, let $\boldsymbol{\Theta}_{\ell p}\in \mathbb{R}^{F_{\ell-1}\times F_\ell}$ denote the filter coefficient matrix at layer $\ell$. The output of layer $\ell$ in a GNN is then recursively given by
\begin{align}
  \label{eq: gnn}
  \mathbf{X}_\ell = \sigma \left( \sum_{p=0}^{P-1} \mathbf{S}^p \mathbf{X}_{\ell-1} \boldsymbol{\Theta}_{\ell p} \right).
\end{align}

Henceforth assume, for simplicity, that the input and output features are one-dimensional ($F_0 = F_L = 1$), while the hidden layers have dimension $F_1 = \dots = F_{L-1} = F$. For a given input vector $\mathbf{x} \in \mathbb{R}^N$, we set $\mathbf{X}_0 = \mathbf{x}$ and define the overall GNN mapping as $u_{\boldsymbol{\theta}}(\mathbf{x}; \mathbf{S})  \coloneq \mathbf{X}_L$, representing the outcome of applying (\ref{eq: gnn}) sequentially. The GNN is parameterized by a tensor of learnable filter coefficients $\boldsymbol{\theta}:=\{\boldsymbol{\Theta}_{\ell p}\}_{\ell,p}$ and a fixed GSO $\mathbf{S}$. The latter endows the GNN with useful inductive biases about the signal's relational structure encoded in $\mathcal{G}$.  

%
%

\subsection{Stability of Graph Neural Networks}
\label{sec: stability-of-gnn}
Let $\mathcal{P}= \{\mathbf{P} \in \{0,1\}^{N \times N}: \mathbf{P}\mathbf{1} = \mathbf{1}, \mathbf{P}^\top\mathbf{1} = \mathbf{1}\}$ be the set of permutation matrices, where $\mathbf{1}\in \mathbb{R}^N$ is the all-ones vector. A fundamental property of GNNs constructed via (\ref{eq: gnn}) is permutation equivariance. Namely, for any permutation matrix $\mathbf{P} \in \mathcal{P}$, if we define the permuted GSO $\hat{\mathbf{S}} = \mathbf{P}^\top \mathbf{S} \mathbf{P}$ and the permuted input signal $\hat{\mathbf{x}} = \mathbf{P}^\top \mathbf{x}$, the GNN output satisfies
\begin{align}
\label{eq: GNN-equivariance}
  u_{\boldsymbol{\theta}}(\hat{\mathbf{x}}; \hat{\mathbf{S}}) = \mathbf{P}^\top u_{\boldsymbol{\theta}}(\mathbf{x}; \mathbf{S}).
\end{align}

Beyond exact permutations, it is also prudent to study how the GNN behaves under structural noise. To this end, we consider the relative perturbation model from \cite{Gama_2020}, where a perturbed GSO $\tilde{\mathbf{S}}$ is related to the nominal GSO $\mathbf{S}$ modulo permutation. Specifically, we evaluate the symmetric error matrix $\mathbf{E}\in\mathcal{S}$ at the permutation $\mathbf{P}_0 \in \mathcal{P}$ that affords the smallest error norm, i.e., 
\begin{equation}
\label{eq: perturbation}
\begin{aligned}
\{\mathbf{E}^\star, \mathbf{P}_0\} &= \argmin_{\mathbf{E}\in\mathcal{S}, \mathbf{P}\in\mathcal{P}} \|\mathbf{E}\|_2\\
\text{s.t.} \quad \mathbf{P}^{\top}\tilde{\mathbf{S}}\mathbf{P} &= \mathbf{S} + \tfrac{1}{2}(\mathbf{E}\mathbf{S}+\mathbf{S}\mathbf{E}).
\end{aligned}
\end{equation}
Unlike absolute perturbations where edge corruptions are uniform, this relative perturbation model ensures that regions of $\mathcal{G}$ with weaker connectivity experience proportionally smaller changes compared to others with stronger links; see also \cite[Sec. IV-A]{luana_2021} for a detailed discussion further justifying the practical relevance of this widely adopted model.

We assume that the pointwise non-linearity $\sigma(\cdot)$ is normalized Lipschitz, i.e., $|\sigma(b) - \sigma(a)| \leq |b-a|$. About the graph filter, we require that the scalar frequency response $h(\lambda)$ is integral Lipschitz with constant $C$ (see \cite[Def. 4]{Gama_2020}) and it is uniformly bounded, i.e., $\|\mathbf{H}(\mathbf{\Lambda})\|_2 \leq B$. Then, GNN outputs are stable, meaning for $\|\mathbf{E}^\star\|_2 \leq \varepsilon$ it holds (see \cite[Thm. 4]{Gama_2020})
\begin{align}
\label{eq: gnn-stability}
  \| \mathbf{P}_0^\top u_{\boldsymbol{\theta}}(\mathbf{x}; \tilde{\mathbf{S}})
  - u_{\boldsymbol{\theta}}(\mathbf{P}_0^\top \mathbf{x}; \mathbf{S}) \|
  \leq (\Gamma \varepsilon + \mathcal{O}(\varepsilon^2)) \|\mathbf{x}\|,
\end{align}
for all $\mathbf{x} \in \mathbb{R}^N$, where $\Gamma := C(1 + \delta \sqrt{N}) L (BF)^{L-1}$, and $\delta\leq 8$ represents the eigenvector misalignment between $\mathbf{E}^{\star}$ and $\mathbf{S}$~\cite[Thm. 1]{Gama_2020}. The quantity $\Gamma$ elucidates the effect of GNN depth $L$ and width $F$ on stability, as well as graph filter spectral response summaries $C$ and $B$ that also play a role.

We leverage this foundational bound to derive the stability of the GNN-parametrized continuous-time flow we introduce in Section \ref{sec: graph flow models} for graph signal generation. However, we note that our analytical framework is general and naturally extends to any base GNN satisfying a stability condition like $\eqref{eq: gnn-stability}$.

%
%

\subsection{Continuous Normalizing Flows and Flow Matching}
Given a dataset of signals $\mathbf{x}^{(1)}, \dots, \mathbf{x}^{(n)}$ drawn from an unknown data distribution $p_{\textrm{data}}$ over $\mathbb{R}^N$, the goal of continuous-time generative modeling is to construct a dynamical system that maps a simple prior distribution $p_{\textrm{init}}$ to $p_{\textrm{data}}$.

A CNF is described by an ODE for times $t\in[0,1]$,
\begin{align*}
\frac{d}{dt}\mathbf{x}_t = u_t^{\boldsymbol{\theta}}(\mathbf{x}_t), \quad \mathbf{x}_0 \sim p_{\textrm{init}},
\end{align*}
where the time-dependent \emph{vector field} $u_t^{\boldsymbol{\theta}}: \mathbb{R}^N \times [0,1] \to \mathbb{R}^N$ is parameterized by a neural network with learnable weights $\boldsymbol{\theta}$. We let $\Phi_t^{\boldsymbol{\theta}}(\mathbf{x}_0)$ denote the \emph{continuous flow map}, defined as the solution to this initial value problem at time $t$ evaluated from $\mathbf{x}_0$. The objective is to train the vector field such that the distribution of the endpoint $\mathbf{x}_1$ closely approximates $p_{\textrm{data}}$.

Let $p_t$ be a target \emph{probability density path} satisfying $p_0 = p_{\textrm{init}}$ and $p_1 = p_{\textrm{data}}$, and let $u_t^{\text{target}}:\mathbb{R}^N \times [0,1] \to \mathbb{R}^N$ be the underlying target vector field that generates this path. The flow matching (FM) loss is defined as
\begin{align*}
  \mathcal{L}_{\textrm{FM}}(\boldsymbol{\theta}) = \mathbb{E}_{\begin{subarray}{l} t\sim\mathcal{U}(0,1) \\ \mathbf{x}\sim p_t \end{subarray}} \left[ \|u_t^{\boldsymbol{\theta}} (\mathbf{x}) - u_t^{\text{target}}(\mathbf{x})\|^2 \right],
\end{align*}
where $\mathcal{U}(0,1)$ stands for the uniform distribution in $(0,1)$.

In practice, both the marginal probability path $p_t$ and the target vector field $u_t^{\text{target}}$ are intractable. To circumvent this, \cite{lipman2022flow} introduces conditional flow matching (CFM). Instead of targeting the marginal vector field directly, we define a conditional probability path $p_t(\cdot|\mathbf{z})$ for each data point $\mathbf{z} \sim p_{\textrm{data}}$. This path satisfies $p_0(\cdot|\mathbf{z}) = p_{\textrm{init}}$, while the endpoint $p_1(\cdot|\mathbf{z})$ is a distribution concentrated around the data point $\mathbf{z}$, such as the multivariate Gaussian $\mathcal{N}(\mathbf{x}; \mathbf{z}, \sigma^2 \mathbf{I})$ for a sufficiently small $\sigma$, where $\mathbf{I}$ denotes the $N\times N$ identity matrix.

If $u_t^{\text{target}}(\cdot|\mathbf{z}): \mathbb{R}^N \times [0,1] \to \mathbb{R}^N$ is the corresponding conditional vector field that generates $p_t(\cdot|\mathbf{z})$, the CFM loss is 
\begin{align}
\label{eq: CFM}
  \mathcal{L}_{\textrm{CFM}}(\boldsymbol{\theta}) = \mathbb{E}_{\begin{subarray}{l} t\sim\mathcal{U}(0,1) \\ \mathbf{z}\sim p_{\textrm{data}} \\ \mathbf{x} \sim p_t(\cdot|\mathbf{z}) \end{subarray}} \left[ \|u_t^{\boldsymbol{\theta}} (\mathbf{x}) - u_t^{\text{target}}(\mathbf{x}|\mathbf{z})\|^2 \right].
\end{align}
Crucially, it follows that $\mathcal{L}_{\textrm{CFM}}(\boldsymbol{\theta})$ and $\mathcal{L}_{\textrm{FM}}(\boldsymbol{\theta})$ are equal up to a constant independent of $\boldsymbol{\theta}$ \cite{lipman2022flow}, allowing us to train the model purely on the conditional paths. For example, a standard choice is the optimal transport path $p_t(\mathbf{x}|\mathbf{z}) = \mathcal{N}(\mathbf{x}; t\mathbf{z}, (1-t)^2 \mathbf{I})$.

%
%
\subsection{Objective and Scope}
\label{sec: problem statement}
Our objective is to learn the distribution $p_{\textrm{data}}$ of signals supported on the nodes of a graph $\mathcal{G}$, starting from a simple prior $p_{\textrm{init}}$. To achieve this, we design a CNF model where the time-dependent vector field $u_t^{\boldsymbol{\theta}}$ is parameterized by a GNN. By embedding the graph topology into the generative dynamics, we prove that the resulting flow is permutation equivariant and derive explicit bounds on its stability under structural errrors adhering to the relative perturbation model in \eqref{eq: perturbation}.
%
%

\section{Graph Flow Models}
\label{sec: graph flow models}

To construct a generative flow model for a given $\mathcal{G}$ with GSO $\mathbf{S}$, we define a time-dependent vector field $u_t^{\boldsymbol{\theta}}(\cdot; \mathbf{S}): \mathbb{R}^N \times [0,1] \to \mathbb{R}^{N}$. Since the data we aim to generate are signals supported on a graph, it is natural to parameterize this continuous vector field using a GNN. Doing so allows the flow dynamics to inherently leverage the structural inductive biases of the underlying graph topology. Because a GNN natively processes only graph signals, we introduce a continuous temporal conditioning function $g : \mathbb{R}^N \times [0,1] \to \mathbb{R}^{N \times d}$ defined by the mapping $(\mathbf{x}, t) \mapsto g(\mathbf{x}, t)$ to incorporate the time variable $t \in [0,1]$. We require $g$ to be permutation equivariant, meaning that for all permutation matrices $\mathbf{P} \in \mathcal{P}$, it holds that
\begin{align}
\label{eq: temporal-conditioning-equivariance}
  g(\hat{\mathbf{x}}, t) = \mathbf{P}^\top g(\mathbf{x}, t).
\end{align}
This condition is naturally satisfied by the standard time-embedding techniques commonly used in generative models \cite{ song2021scorebasedgenerativemodelingstochastic, nichol2021improved}.
For instance, given a time embedding $\textbf{emb}: \mathbb{R} \to \mathbb{R}^{d-1}$, $g$ can be implemented via concatenation as $g(\mathbf{x}, t) = [\mathbf{x}, \mathbf{1}\textbf{emb}(t)^\top] \in \mathbb{R}^{N \times d}$; or via addition as $g(\mathbf{x}, t) = \mathbf{x} + \mathbf{1}\text{emb}(t) \in \mathbb{R}^{N}$ using a scalar $\text{emb}: \mathbb{R} \to \mathbb{R}$. Because an identical time embedding is applied across all nodes, the operation remains entirely independent of the node indexing and \eqref{eq: temporal-conditioning-equivariance} holds. Without loss of generality, to simplify the notational burden in the subsequent theoretical analysis, we will henceforth assume $d=1$, such that $g(\mathbf{x}, t) \in \mathbb{R}^N$. If a multi-dimensional mapping is used in practice, the stability bounds derived in Sections \ref{sec: stability-cts} and \ref{sec: stability-discrete} naturally extend across the feature dimensions via the triangle inequality.

The vector field is then defined by evaluating a GNN on the time-augmented graph signal $g(\mathbf{x}, t)$, namely
\begin{align}
\label{eq: VF}
  u_t^{\boldsymbol{\theta}}(\mathbf{x}; \mathbf{S})  \coloneq u_{\boldsymbol{\theta}} \big(g(\mathbf{x},t); \mathbf{S}\big).
\end{align}
Given an initial condition $\mathbf{x}_0 \sim p_{\textrm{init}}$, the continuous-time dynamics of our generative model are governed by the ODE
\begin{align}
  \label{eq: ODE}
  \frac{d}{dt}\mathbf{x}_t = u_t^{\boldsymbol{\theta}}(\mathbf{x}_t; \mathbf{S}).
\end{align}
We denote by $\Phi_t^{\boldsymbol{\theta}} (\mathbf{x}_0; \mathbf{S})$ the associated graph-aware flow map. 

To guarantee the existence and uniqueness of $\Phi_t^{\boldsymbol{\theta}}$ via the Picard-Lindelöf theorem, we assume the vector field is Lipschitz continuous in the state variable. This is a standard property of neural networks, including GNNs (see Proposition \ref{prop: lipschitz-bound}). Specifically, we assume there exists a constant $M > 0$ such that for all graph signals $\mathbf{x}, \mathbf{y} \in \mathbb{R}^N$ and all $t \in [0,1]$, 
\begin{align}
\label{eq: lipschitz}
  \|u_t^{\boldsymbol{\theta}} (\mathbf{x}; \mathbf{S}) - u_t^{\boldsymbol{\theta}}(\mathbf{y}; \mathbf{S})\| \leq M \|\mathbf{x}-\mathbf{y}\|.
\end{align}

%
%

\subsection{Permutation Equivariance of the Continuous Flow}

Having defined our GNN-parametrized flow model, the foremost property to verify is permutation equivariance. We first establish the permutation equivariance of the vector field.

\begin{definition}[Permutation equivariant vector field]
\label{def: perm-equiv}
We say a vector field $u_t^{\boldsymbol{\theta}}$ is permutation equivariant if it satisfies
\begin{align*}
  u_t^{\boldsymbol{\theta}}(\hat{\mathbf{x}}; \hat{\mathbf{S}}) = \mathbf{P}^\top u_t^{\boldsymbol{\theta}}(\mathbf{x}; \mathbf{S}),
\end{align*}
for all $\mathbf{P} \in \mathcal{P}$, $\mathbf{x} \in \mathbb{R}^N$, and $t \in [0,1]$.
\end{definition}

By the equivariance of the temporal conditioning $g$ in \eqref{eq: temporal-conditioning-equivariance} and the base GNN operator $u_{\boldsymbol{\theta}}$ \eqref{eq: GNN-equivariance}, we can explicitly verify this property holds for the vector field \eqref{eq: VF} as well. For any permutation matrix $\mathbf{P} \in \mathcal{P}$, evaluating the vector field on the permuted state $\hat{\mathbf{x}} = \mathbf{P}^\top \mathbf{x}$ and graph $\hat{\mathbf{S}} = \mathbf{P}^\top \mathbf{S} \mathbf{P}$ yields
\begin{equation}
\begin{aligned}
\label{eq: VF-equivariance}
  u_t^{\boldsymbol{\theta}}(\hat{\mathbf{x}}; \hat{\mathbf{S}}) &{}= u_{\boldsymbol{\theta}}\big(g(\hat{\mathbf{x}}, t); \hat{\mathbf{S}}\big) = u_{\boldsymbol{\theta}}\big(\mathbf{P}^\top g(\mathbf{x}, t); \hat{\mathbf{S}}\big) \\
  &{}= \mathbf{P}^\top u_{\boldsymbol{\theta}}\big(g(\mathbf{x}, t); \mathbf{S}\big) = \mathbf{P}^\top u_t^{\boldsymbol{\theta}}(\mathbf{x}; \mathbf{S}).
\end{aligned}
\end{equation}
Thus, the vector field satisfies Definition \ref{def: perm-equiv}.

Ultimately, our goal is to ensure that the generated signals themselves exhibit this structural symmetry. The following proposition states that indeed the resulting flow map $\Phi_t^{\boldsymbol{\theta}} (\mathbf{x}_0; \mathbf{S})$ inherits the permutation equivariance of the vector field.

\begin{proposition}[Permutation equivariance of the flow]
\label{prop: flow-equivariance}
Assume the vector field $u_t^{\boldsymbol{\theta}}$ is permutation equivariant and Lipschitz continuous. Then, the associated flow map satisfies
\begin{equation}
\label{eq:flow-equivariance}
  \Phi_t^{\boldsymbol{\theta}}(\hat{\mathbf{x}}_0;\hat{\mathbf{S}}) = \mathbf{P}^\top \Phi_t^{\boldsymbol{\theta}}(\mathbf{x}_0;\mathbf{S}),
\end{equation}
for all $t \in [0,1]$ and $\mathbf{x}_0 \in \mathbb{R}^N$.
\end{proposition} 

\noindent\textit{Proof.} See Appendix \ref{proof: flow-equivariance}.

In particular, evaluating \eqref{eq:flow-equivariance} at $t=1$ yields the equivariance of the generated graph signal sample. If we consider the initial state as a random vector $\mathbf{x}_0 \sim p_{\textrm{init}}$, Proposition \ref{prop: flow-equivariance} implies that the random vectors $\Phi_t^{\boldsymbol{\theta}}(\hat{\mathbf{x}}_0;\hat{\mathbf{S}})$ and $\mathbf{P}^\top \Phi_t^{\boldsymbol{\theta}}(\mathbf{x}_0;\mathbf{S})$ are equal almost surely. Therefore, they are equal in distribution for any $p_{\textrm{init}}$, which we denote as
\begin{align*}
  \Phi_t^{\boldsymbol{\theta}}(\hat{\mathbf{x}}_0;\hat{\mathbf{S}}) \overset{D}{=} \mathbf{P}^\top \Phi_t^{\boldsymbol{\theta}}(\mathbf{x}_0;\mathbf{S}).
\end{align*}

%
%

\subsection{Stability of the Continuous Flow}
\label{sec: stability-cts}
With the permutation equivariance of the flow established in Proposition \ref{prop: flow-equivariance}, we can now move on to assess its structural stability. We reiterate that because node indexing is often arbitrary, comparing outputs on a nominal GSO $\mathbf{S}$ and a perturbed one $\tilde{\mathbf{S}}$ must be carried out modulo permutation. 

Recall that \eqref{eq: gnn-stability} asserts the stability of the base GNN, $u_{\boldsymbol{\theta}}$, under relative perturbations. To determine whether this desired property translates to the flow, we first bound the perturbation error of the vector field $u_t^{\boldsymbol{\theta}}(\mathbf{x}; \mathbf{S}) = u_{\boldsymbol{\theta}}(g(\mathbf{x},t); \mathbf{S})$ evaluated at an arbitrary state. To this end, note that (\ref{eq: gnn-stability}) holds for all $\mathbf{x} \in \mathbb{R}^N$. Thus, we can plug the signal $g(\mathbf{x},t)$ into it, so that the perturbation error of our parameterized vector field is immediately bounded by
\begin{align}
\label{eq: VF-stability}
\hspace{-7pt}\| \mathbf{P}_0^\top  u_t^{\boldsymbol{\theta}}(\mathbf{x}; \tilde{\mathbf{S}}) -  u_t^{\boldsymbol{\theta}}(\mathbf{P}_0^\top \mathbf{x}; \mathbf{S}) \| \leq  (\Gamma \varepsilon + \mathcal{O}(\varepsilon^{2}))\| g(\mathbf{x},t) \|, 
\end{align}
for all $\mathbf{x} \in \mathbb{R}^N$ and $t\in [0,1]$.

Because the vector field is Lipschitz continuous [cf. \eqref{eq: lipschitz}], the flow is continuous with respect to time and initial conditions. Let $\mathbf{y}_t^{\mathbf{P}}  \coloneq \Phi_t^{\boldsymbol{\theta}}(\mathbf{P}^\top \mathbf{x}_0; \mathbf{S})$ denote the trajectory starting from a permuted initial state. Since $g$ is continuous and the set of permutation matrices $\mathcal{P}$ is finite, the signal $g(\mathbf{y}_t^{\mathbf{P}}, t)$ is strictly bounded across all possible permuted trajectories. This guarantees the existence of a finite constant 
$C_g(\mathbf{x}_0)  \coloneq \max_{\mathbf{P} \in \mathcal{P}} \sup_{\tau \in [0,1]} \|g(\mathbf{y}_\tau^{\mathbf{P}}, \tau)\| < \infty$.

Since the vector field is Lipschitz continuous with constant $M$, it is also one-sided Lipschitz with a global constant $m \leq M$. We define an integrable time-dependent one-sided Lipschitz coefficient $m_t \in L^1([0,1])$ such that
\begin{align}
  \label{eq: one-sided lipschitz}
  \langle u_t^{\boldsymbol{\theta}}(\mathbf{x};\mathbf{S}) - u_t^{\boldsymbol{\theta}} (\mathbf{y};\mathbf{S}), \mathbf{x}-\mathbf{y} \rangle \leq m_t \|\mathbf{x}-\mathbf{y}\|^2,
\end{align}
for all $\mathbf{x},\mathbf{y} \in \mathbb{R}^N$. Equipped with this one-sided Lipschitz condition, we can formalize the stability of the flow model.

\begin{theorem}[Stability of the flow model]
\label{thm: flow-stability}
Assuming the vector field $u_t^{\boldsymbol{\theta}}$ is one-sided Lipschitz with coefficient $m_t$, and the base GNN, $u_{\boldsymbol{\theta}}$, satisfies the stability conditions yielding \eqref{eq: gnn-stability}, then it holds that
\begin{equation}
\begin{split}
  \min_{\mathbf{P} \in \mathcal{P}} \|\mathbf{P}^\top \Phi_t^{\boldsymbol{\theta}}( \mathbf{x}_0; \tilde{\mathbf{S}}) &- \Phi_t^{\boldsymbol{\theta}}(\mathbf{P}^\top \mathbf{x}_0; \mathbf{S}) \| \\
  &\leq \Omega_tC_g(\mathbf{x}_0) \big(\Gamma \varepsilon+ \mathcal{O}(\varepsilon^2)\big),
\end{split}\label{eq:stability_flow_model}
\end{equation}
for all $t \in [0,1]$ and $\mathbf{x}_0 \in \mathbb{R}^N$, where
\begin{align}
  \Omega_t & \coloneq \int_0^t \exp\left({\int_\tau^t m_s ds}\right) d\tau, \label{eq:omega_t}\\
  C_g(\mathbf{x}_0) & \coloneq \max_{\mathbf{P} \in \mathcal{P}} \sup_{\tau \in [0,1]} \|g(\Phi_\tau^{\boldsymbol{\theta}}(\mathbf{P}^\top \mathbf{x}_0; \mathbf{S}), \tau)\|.\nonumber
\end{align}
\end{theorem}

\noindent\textit{Proof.} See Appendix \ref{proof: flow-stability}.

To obtain more interpretable closed-form expressions, we can evaluate the integral $\Omega_t$ using the global constants $m$ and $M$ introduced previously.

\begin{corollary}[Simplified bounds of the flow model]
\label{cor: bounds}
Under the assumptions of Theorem \ref{thm: flow-stability}, substituting the global Lipschitz constants simplifies the integral $\Omega_t$ in \eqref{eq:omega_t} as follows:
\begin{enumerate}[label=(\roman*)]
  \item For a global one-sided Lipschitz constant $m$, then $\Omega_t = \frac{e^{mt}-1}{m}$.
  \item For the standard Lipschitz constant $M$, then $\Omega_t = \frac{e^{Mt}-1}{M}$.
\end{enumerate}
\end{corollary}

Furthermore, evaluating the bound \eqref{eq:stability_flow_model} of Theorem \ref{thm: flow-stability} at time $t=1$ yields the stability of the generated sample:
\begin{equation*}
\begin{split}
  \min_{\mathbf{P} \in \mathcal{P}} \|\mathbf{P}^\top \Phi_1^{\boldsymbol{\theta}}( \mathbf{x}_0; \tilde{\mathbf{S}}) &- \Phi_1^{\boldsymbol{\theta}}(\mathbf{P}^\top \mathbf{x}_0; \mathbf{S}) \| \\
  &\leq \Omega_1 C_g(\mathbf{x}_0) \big(\Gamma \varepsilon+ \mathcal{O}(\varepsilon^2)\big),
\end{split}
\end{equation*}
for all $\mathbf{x}_0 \in \mathbb{R}^N$.

So far, our results characterize the stability of individual trajectories starting from a fixed, deterministic initial state $\mathbf{x}_0$. However, CNFs ultimately model probability distributions. By treating the initial condition as a random vector $\mathbf{x}_0 \sim p_{\textrm{init}}$, we can broaden these pointwise results to bound the 2-Wasserstein distance \cite{villani2009optimal} between the generated distributions.

\begin{corollary}[Wasserstein stability of the flow model]
\label{cor: wasserstein-stability}
Under the assumptions of Theorem \ref{thm: flow-stability}, let the initial state be a random vector $\mathbf{x}_0 \sim p_{\textrm{init}}$. Let $\mathbf{P}^\top \Phi_t^{\boldsymbol{\theta}}(\mathbf{x}_0;\tilde{\mathbf{S}}) \sim \tilde{\pi}_t^{\mathbf{P}}$ and $\Phi_t^{\boldsymbol{\theta}}(\mathbf{P}^\top \mathbf{x}_0;\mathbf{S}) \sim \pi_t^{\mathbf{P}}$. Then, it holds that 
\begin{equation}\label{eq:bound_cor_2}
  \min_{\mathbf{P}\in\mathcal{P}} W_2( \tilde{\pi}_t^{\mathbf{P}}, \pi_t^{\mathbf{P}} ) \leq \Omega_t C_g' \big(\Gamma \varepsilon + \mathcal{O}(\varepsilon^2)\big),
\end{equation}
for all $t \in [0,1]$, where $C_g' = \sqrt{\mathbb{E}_{\mathbf{x}_0}\big[C_g^2(\mathbf{x}_0)\big]}$.
\end{corollary}

\noindent\textit{Proof.} See Appendix \ref{proof: wasserstein-stability}.

In summary, these stability results are governed by three terms: (i) $\Gamma$, which captures the inherent stability of the base GNN; (ii) $\Omega_1$, which is exponential in the Lipschitz constant and dictates how the continuous dynamics amplify errors; and (iii) $C_g(\mathbf{x}_0)$, the trajectory supremum across all initial condition permutations. Notably, the expectation $C_g'$ becomes independent of $\mathbf{P}$ when $p_{\textrm{init}}(\mathbf{x}) = \mathcal{N}(\mathbf{x};\mathbf{0}, \mathbf{I})$, since the initial and permuted pushforward distributions are identical.

%
%

\section{Discrete Graph Signal Samplers}
\label{sec: discrete samplers}

The flow map $\Phi_t^{\boldsymbol{\theta}}$ associated with the ODE in \eqref{eq: ODE} defines the continuous-time dynamics of our generative model. Even though this provides the exact model for transporting the prior $p_{\textrm{init}}$ to the desired data distribution $p_{\textrm{data}}$, the resulting trajectories cannot be computed analytically in practice. To actually obtain a generated sample, we must approximate the continuous flow using discrete samplers, such as the Euler and Heun methods \cite{butcher2016numerical}. We discretize the time interval $[0,1]$ into $K$ steps of size $h = 1/K$, corresponding to times $t_k = kh$. To streamline notation, we will henceforth use the iteration index $k$ interchangeably with the time $t_k$, writing the vector field evaluated at step $k$ simply as $u_k^{\boldsymbol{\theta}}(\cdot; \mathbf{S}) \coloneq u_{t_k}^{\boldsymbol{\theta}}(\cdot; \mathbf{S})$. Under this discretization, the final iterate $\mathbf{x}_K$ approximates the terminal sample of the continuous flow, $\Phi_1^{\boldsymbol{\theta}}(\mathbf{x}_0; \mathbf{S})$.

Formally, given a GSO $\mathbf{S}$, a deterministic $K$-step sampler is defined by a sequence of update maps $\{T_k^{\boldsymbol{\theta}}(\cdot;\mathbf{S})\}_{k=0}^{K-1}$ that generates a trajectory $\{\mathbf{x}_k\}_{k=0}^K$ according to $\mathbf{x}_{k+1} = T_k^{\boldsymbol{\theta}}(\mathbf{x}_k;\mathbf{S})$, for $k=0,\dots,K-1$. From this, we define the associated \emph{discrete flow map} $\Psi_k^{\boldsymbol{\theta}}(\mathbf{x}_0; \mathbf{S}): \mathbb{R}^N \to \mathbb{R}^N$ at step $k \in \{0, \dots, K\}$ recursively as:
\begin{align}
\label{eq: discrete-flow-map}
    \Psi_0^{\boldsymbol{\theta}}(\mathbf{x}_0; \mathbf{S}) &\coloneqq \mathbf{x}_0, \nonumber \\
    \Psi_k^{\boldsymbol{\theta}}(\mathbf{x}_0; \mathbf{S}) &\coloneqq T_{k-1}^{\boldsymbol{\theta}}\big(\Psi_{k-1}(\mathbf{x}_0; \mathbf{S}); \mathbf{S}\big), \quad \text{for } k \geq 1.
\end{align}

Specific numerical methods provide distinct instantiations of the maps $T_k^{\boldsymbol{\theta}}$ in terms of the parameterized graph-aware vector field $u_k^{\boldsymbol{\theta}}$. For instance, the Euler sampler yields
\begin{align}
\label{eq: euler-update}
        T_k^{\boldsymbol{\theta}}(\mathbf{x};\mathbf{S}) = \mathbf{x} + h\,u^{\boldsymbol{\theta}}_{k}(\mathbf{x};\mathbf{S}),
\end{align}
while the Heun sampler is defined by the update map
\begin{align}
\label{eq: heun-update}
    T_k^{\boldsymbol{\theta}}(\mathbf{x};\mathbf{S}) = \mathbf{x} + \frac{h}{2}\big(u_k^{\boldsymbol{\theta}} (\mathbf{x};\mathbf{S}) + u_{k+1}^{\boldsymbol{\theta}}(\mathbf{x}+hu_k^{\boldsymbol{\theta}}(\mathbf{x};\mathbf{S});\mathbf{S})\big).
\end{align}

%
%

\subsection{Permutation Equivariance of the Discrete Flow}

As for the continuous-time model, the foremost property to verify for any viable numerical approximation is that it preserves the structural symmetries in graph $\mathcal{G}$ with GSO $\mathbf{S}$.

\begin{definition}[Permutation equivariant sampler]
We say a deterministic $K$-step sampler is permutation equivariant if its update maps satisfy
\begin{align}
\label{eq: sampler-equivariance}
  T_k^{\boldsymbol{\theta}}(\hat{\mathbf{x}}; \hat{\mathbf{S}}) = \mathbf{P}^\top T_k^{\boldsymbol{\theta}}(\mathbf{x};\mathbf{S}),
\end{align}
for all $k \in \{0, \dots, K-1\}$, $\mathbf{x} \in \mathbb{R}^N$, and $\mathbf{P} \in \mathcal{P}$.
\end{definition}

\begin{proposition}[Permutation equivariance of the discrete flow]
\label{prop: discrete-flow-equivariance}
Assume the underlying sampler $\{T_k^{\boldsymbol{\theta}}(\cdot;\mathbf{S})\}_{k=0}^{K-1}$ is permutation equivariant. Then, the discrete flow map in \eqref{eq: discrete-flow-map} satisfies
\begin{align*}
  \Psi_k^{\boldsymbol{\theta}}(\hat{\mathbf{x}}_0; \hat{\mathbf{S}}) = \mathbf{P}^\top \Psi_k^{\boldsymbol{\theta}}(\mathbf{x}_0; \mathbf{S}),
\end{align*}
for all $k \in \{0, \dots, K\}$, $\mathbf{x}_0 \in \mathbb{R}^N$ and $\mathbf{P} \in \mathcal{P}$.
\end{proposition}

\noindent\textit{Proof.} See Appendix \ref{proof: discrete-flow-equivariance}.

In particular, Proposition \ref{prop: discrete-flow-equivariance} guarantees the equivariance of the generated terminal sample, $\Psi_K^{\boldsymbol{\theta}}(\hat{\mathbf{x}}_0; \hat{\mathbf{S}}) = \mathbf{P}^\top \Psi_K^{\boldsymbol{\theta}}(\mathbf{x}_0;\mathbf{S})$. This condition is naturally satisfied by many widely used discrete samplers. Notably, if the underlying vector field $u_k^{\boldsymbol{\theta}}$ is permutation equivariant (cf. Definition \ref{def: perm-equiv}), the Euler and Heun samplers defined in \eqref{eq: euler-update} and \eqref{eq: heun-update} are permutation equivariant as well. And by Proposition \ref{prop: discrete-flow-equivariance}, the discrete flow maps induced by both methods are permutation equivariant. The full details to establish the equivariance of these samplers are provided in Appendices \ref{proof: euler-equivariance} and \ref{proof: heun-equivariance}.

\subsection{Stability of the Discrete Flow}
\label{sec: stability-discrete}
Building on the continuous case in Section \ref{sec: stability-cts}, our goal is to bound the deviation between a trajectory generated on the perturbed GSO $\tilde{\mathbf{S}}$, where $\tilde{\mathbf{S}}$ is defined as in \eqref{eq: perturbation}, and the appropriately permuted trajectory on the nominal GSO $\mathbf{S}$. Arguably, this is the most important stability consideration when it comes to generating graph signals in practice. By expanding the recursive definition of the discrete flow, the error at any step $k+1$ can be decomposed using the triangle inequality into two distinct sources of error. This motivates two key characterizations of the numerical method: (i) how robust the sampler's update map is to graph perturbations (graph stability); and (ii) how sensitive it is to input variations (state stability).

\begin{definition}[Graph stable sampler]
\label{def: graph-stable}
We say a permutation equivariant sampler $\{T_k^{\boldsymbol{\theta}}(\cdot; \mathbf{S})\}_{k=0}^{K-1}$ is graph stable if there exists a permutation invariant bounding function $\beta: \mathbb{R}^N \times [0,1] \to \mathbb{R}_{\geq 0}$ (i.e., $\beta(\mathbf{P}^\top \mathbf{x}, t) = \beta(\mathbf{x}, t)$ for all $\mathbf{P} \in \mathcal{P}$) and a constant $\Gamma>0$, such that
\begin{align*}
  \|\mathbf{P}_0^\top T_k^{\boldsymbol{\theta}}(\mathbf{x};\tilde{\mathbf{S}})-T_k^{\boldsymbol{\theta}}(\mathbf{P}_0^\top \mathbf{x};\mathbf{S})\| \le \beta(\mathbf{x}, t_k) \big(\Gamma\varepsilon + \mathcal{O}(\varepsilon^2)\big),
\end{align*}
for all $\mathbf{x}\in \mathbb{R}^N$ and $k \in \{0,\dots, K-1\}$.
\end{definition}

\begin{definition}[State stable sampler]
\label{def: state-stable}
We say a sampler $\{T_k^{\boldsymbol{\theta}}(\cdot; \mathbf{S})\}_{k=0}^{K-1}$ is state stable if there exists a constant $\alpha \ge 0$, such that
\begin{align*}
  \|T_k^{\boldsymbol{\theta}}(\mathbf{x};\mathbf{S})-T_k^{\boldsymbol{\theta}}(\mathbf{y};\mathbf{S})\| \leq \alpha \|\mathbf{x}-\mathbf{y}\|,
\end{align*}
for all $\mathbf{x},\mathbf{y} \in \mathbb{R}^N$ and $k \in \{0,\dots, K-1\}$.
\end{definition}

Equipped with these two characterizations, we can establish a general stability bound for the resulting discrete flow map by tracking the accumulated error across the $K$ integration steps.

\begin{theorem}[Stability of the discrete flow] 
\label{thm: discrete-flow-stability}
Assuming the underlying permutation equivariant sampler $\{T_k^{\boldsymbol{\theta}}(\cdot; \mathbf{S})\}_{k=0}^{K-1}$ is graph stable with bounding function $\beta$, and state stable with constant $\alpha$. Then, it holds that
\begin{equation*}
  \begin{aligned}
    \min_{\mathbf{P} \in \mathcal{P}} \|\mathbf{P}^\top \Psi_k^{\boldsymbol{\theta}}( \mathbf{x}_0; \tilde{\mathbf{S}}) 
    &- \Psi_k^{\boldsymbol{\theta}}(\mathbf{P}^\top \mathbf{x}_0; \mathbf{S}) \| \\
    & \leq \Upsilon_k(\alpha)\,\beta_{\max} \big(\Gamma\varepsilon + \mathcal{O}(\varepsilon^2)\big),
  \end{aligned}
\end{equation*}
for all $k \in \{1, \dots, K\}$ and $\mathbf{x}_0 \in \mathbb{R}^N$, where
\begin{align*}
  \Upsilon_k(\alpha) & \coloneq
  \begin{cases}
  \frac{1-\alpha^k}{1-\alpha} & \text{if } \alpha \neq 1, \\
  k & \text{if } \alpha = 1,
  \end{cases} \\
  \beta_{\max} & \coloneq \max_{\mathbf{P} \in \mathcal{P}} \max_{j \in \{0, \dots, K-1\}} \beta(\mathbf{y}_j^{\mathbf{P}}, t_j), \\
  \mathbf{y}_j^{\mathbf{P}} & \coloneq \Psi_j^{\boldsymbol{\theta}}(\mathbf{P}^\top \mathbf{x}_0; \mathbf{S}).
\end{align*}
\end{theorem}

\noindent\textit{Proof.} See Appendix \ref{proof: discrete-flow-stability}.

As established previously, both the Euler and Heun methods are permutation equivariant, allowing us to directly apply this general stability framework. By evaluating these specific numerical methods under the standard assumption of Lipschitz continuity, we obtain the following explicit stability bounds.

\begin{corollary}[Stability of Euler and Heun samplers]
\label{cor: euler-heun-stability}
Under the assumptions of Theorem \ref{thm: flow-stability}, if the vector field $u_t^{\boldsymbol{\theta}}$ is Lipschitz continuous with constant $M$, then the discrete flow maps induced by the respective samplers satisfy the following bounds for all $k \in \{1, \dots, K\}$ and $\mathbf{x}_0 \in \mathbb{R}^N$:
\begin{enumerate}[label=(\roman*)]
  \item \textbf{Euler sampler:}
  \begin{equation*}
  \begin{split}
    \min_{\mathbf{P} \in \mathcal{P}} &\|\mathbf{P}^\top \Psi_k^{\boldsymbol{\theta}}( \mathbf{x}_0; \tilde{\mathbf{S}}) - \Psi_k^{\boldsymbol{\theta}}(\mathbf{P}^\top \mathbf{x}_0; \mathbf{S}) \| \\
    &\leq \frac{(1+ hM)^k -1}{M} C_g(\mathbf{x}_0) \big(\Gamma\varepsilon + \mathcal{O}(\varepsilon^2)\big),
  \end{split}
  \end{equation*}
  where
  \begin{align*}
    C_g(\mathbf{x}_0) & \coloneq \max_{\mathbf{P} \in \mathcal{P}} \max_{j < k} \|g(\mathbf{y}_j^{\mathbf{P}}, t_j)\|.
  \end{align*}
  
  \item \textbf{Heun sampler:}
  \begin{equation*}
  \begin{split}
    \min_{\mathbf{P} \in \mathcal{P}} &\|\mathbf{P}^\top \Psi_k^{\boldsymbol{\theta}}( \mathbf{x}_0; \tilde{\mathbf{S}}) - \Psi_k^{\boldsymbol{\theta}}(\mathbf{P}^\top \mathbf{x}_0; \mathbf{S}) \| \\
    &\leq \frac{\left(1+ hM + \frac{h^2M^2}{2}\right)^k -1}{M} C_g(\mathbf{x}_0) \big(\Gamma\varepsilon + \mathcal{O}(\varepsilon^2)\big),
  \end{split}
  \end{equation*}
  where
  \begin{align*}
    C_g(\mathbf{x}_0)  \coloneq \max_{\mathbf{P} \in \mathcal{P}} \max_{j < k} & \max \Big( \|g(\mathbf{y}_j^{\mathbf{P}}, t_j)\|, \\
    &\|g(\mathbf{y}_j^{\mathbf{P}} + hu_j^{\boldsymbol{\theta}}(\mathbf{y}_j^{\mathbf{P}};\mathbf{S}), t_{j+1})\| \Big).
  \end{align*}
\end{enumerate}
Both bounds are evaluated along the unperturbed trajectories defined by $\mathbf{y}_j^{\mathbf{P}}  \coloneq \Psi_j^{\boldsymbol{\theta}}(\mathbf{P}^\top \mathbf{x}_0; \mathbf{S})$.
\end{corollary}
\noindent \textit{Proof.} See Appendix \ref{proof: euler-heun-stability}.

A few noteworthy observations can be made regarding these discrete stability bounds. First, their structure perfectly mirrors the bounds established in Theorem \ref{thm: flow-stability}, consisting of a trajectory dependent supremum scaled by a time dependent growth factor. Notably, in the limit as the step size $h \to 0$ (and the number of steps $k \to \infty$ such that $kh = t$), the growth factors for both Euler and Heun samplers converge to their continuous-time counterpart [cf. (ii) in Corollary \ref{cor: bounds}]. Second, just as we did for the continuous flow, we can evaluate these discrete bounds at the final integration step $k=K$ (which corresponds to time $t=1$). This yields the explicit stability bound for the generated sample. 

Again, our ultimate robustness objective is to measure the stability of the generated distributions. By treating the initial condition as a random vector $\mathbf{x}_0 \sim p_{\textrm{init}}$, we can lift the pointwise guarantees of our numerical samplers to a Wasserstein stability bound for the discrete flow.

\begin{corollary}[Wasserstein stability of the discrete flow]
\label{cor: discrete-wasserstein-stability}
Under the assumptions of Theorem \ref{thm: discrete-flow-stability}, let the initial state be a random vector $\mathbf{x}_0 \sim p_{\textrm{init}}$. Let the resulting distributions at step $k$ be denoted by $\mathbf{P}^\top \Psi_k^{\boldsymbol{\theta}}(\mathbf{x}_0;\tilde{\mathbf{S}}) \sim \tilde{\pi}_k^{\mathbf{P}}$ and $\Psi_k^{\boldsymbol{\theta}}(\mathbf{P}^\top \mathbf{x}_0;\mathbf{S}) \sim \pi_k^{\mathbf{P}}$. Then, it holds that
\begin{equation}\label{eq:bound_cor_4}
  \min_{\mathbf{P} \in \mathcal{P}} W_2( \tilde{\pi}_k^{\mathbf{P}}, \pi_k^{\mathbf{P}} ) 
  \leq \Upsilon_k(\alpha)\,\beta'_{\max} \big(\Gamma\varepsilon + \mathcal{O}(\varepsilon^2)\big),
\end{equation}
for all $k \in \{1, \dots, K\}$, where $\beta'_{\max}  \coloneq \sqrt{\mathbb{E}_{\mathbf{x}_0}\big[\beta_{\max}^2(\mathbf{x}_0)\big]}$.
\end{corollary}

\noindent\textit{Proof.} See Appendix \ref{proof: discrete-wasserstein-stability}.
%
%

\section{Implications for Training Regularization}
\label{sec: regularization}

Beyond providing theoretical guarantees, the stability bounds derived in the previous sections offer actionable information on how to regularize our generative model to promote robustness. By inspecting these expressions, we can isolate the factors inherent to the data from those governed by our design choices. The graph size $N$ is fixed, and the eigenvector misalignment $\delta$ depends entirely on the perturbed GSO $\tilde{\mathbf{S}}$. On the other hand, the GNN parameters $L$, $F$, $B$, $C$, and $M$ are determined by our architecture and learned filter weights.

The number of layers $L$ (GNN depth) and hidden features $F$ (GNN width) dictate the architectural capacity of the vector field. As evidenced by the factor $(BF)^{L-1}$, increasing depth and width exponentially amplifies instability \cite{Gama_2020}. As expected, this implies the architecture should be kept as simple as possible. While acknowledging this structural trade-off is standard practice, the true potential for principled regularization lies in actively controlling the filter-dependent constants $B$, $C$, and $M$ during training.

Recall that the filter's frequency response is bounded such that $\|\mathbf{H}(\mathbf{\Lambda})\|_2 \leq B$. Expressed in terms of the filter coefficients $\boldsymbol{\Theta}_{\ell p}$, the infimum $B$ that satisfies this condition across all layers $\ell=1,\ldots,L$ is given by
\begin{align}
\label{eq: B}
  B = \max_{\ell} \max_{i \in \{1, \dots, n\}} \left\| \sum_{p=0}^{P-1} \boldsymbol{\Theta}_{\ell p} \lambda_i^p \right\|_2.
\end{align}
Similarly, the integral Lipschitz condition implies that $\left|\lambda \frac{dh(\lambda)}{d\lambda}\right| \leq C$. Expressed in terms of the filter coefficients, the infimum $C$ satisfying this inequality is bounded by
\begin{align}
\label{eq: C}
  C \leq \max_{\ell} \max_{i \in \{1, \dots, n\}} \left\| \sum_{p=0}^{P-1} p \boldsymbol{\Theta}_{\ell p} \lambda_i^p \right\|_2.
\end{align}

Beyond bounding the filter conditions of the base GNN, next we establish an explicit bound for the spatial Lipschitz constant $M$ of the vector field introduced in \eqref{eq: lipschitz}.

\begin{proposition}[Lipschitz constant of the vector field]
\label{prop: lipschitz-bound}
The infimum spatial Lipschitz constant $M$  of the vector field \eqref{eq: VF} is bounded by
\begin{equation}
\label{eq: M}
  M \leq \prod_{\ell=1}^L \max_{i \in \{1, \dots, n\}} \left\| \sum_{p=0}^{P-1} \boldsymbol{\Theta}_{\ell p} \lambda_i^p \right\|_2.
\end{equation}
\end{proposition}

\noindent\textit{Proof.} See Appendix \ref{proof: lipschitz-bound}.

We note that if the temporal conditioning function $g$ decouples the state $\mathbf{x}_t$ and time $t$, for example, via concatenation $g(\mathbf{x}, t) = [\mathbf{x}, \mathbf{1}\textbf{emb}(t)^\top] \in \mathbb{R}^{N \times d}$, the time component does not affect the spatial Lipschitz constant. This follows as the norm of the difference perfectly isolates the state
\begin{equation*}
  \|g(\mathbf{x}, t) - g(\mathbf{y}, t)\|_F = \|\mathbf{x} - \mathbf{y}\|.
\end{equation*}
Consequently, the weights of the first layer operating exclusively on the time embedding do not contribute to the bound in Proposition \ref{prop: lipschitz-bound}.

All in all, the bounds in \eqref{eq: B}, \eqref{eq: C} and \eqref{eq: M}, provide a principled basis to regularize $B$, $C$, and $M$, respectively. A key observation is that these bounds depend directly on the spectrum $\{\lambda_i\}_{i=1}^N$ of the given GSO $\mathbf{S}$, making them fundamentally graph-aware. Furthermore, since the underlying graph topology is fixed, this spectrum only needs to be computed once prior to training. Related works have explored regularization strategies based on similar quantities for static GNNs \cite{arghal2021robustgraphneuralnetworks, cervino2022training}. However, highlighting the specific theoretical results of our work, the derived bounds suggest that the spatial Lipschitz constant $M$ acts as the dominating term since it drives the exponential growth factor $\frac{e^M}{M}$ (see Corollary \ref{cor: bounds}). Consequently, for the scope of this paper, our proposed regularization strategy solely penalizes $M$ to ensure stable generative dynamics across CNF training paradigms, though alternative approaches could be chosen to jointly regularize both the spatial and integral Lipschitz constants.

To demonstrate these ideas, we consider a regularized training objective of the form 
\begin{align}
\label{eq: RFM}
  \mathcal{L}(\boldsymbol{\theta}) = \mathcal{L}_{\textrm{CFM}}(\boldsymbol{\theta}) + \mu \mathcal{R}(\boldsymbol{\theta}),
\end{align}
where the $\mathcal{L}_{CFM}$ is defined as in \eqref{eq: CFM} and the regularization term $\mathcal{R}$ is defined as the bound in Proposition~\ref{prop: lipschitz-bound}, while $\mu>0$ is a tuning hyperparameter.

%
%

\section{Numerical Experiments}
\label{sec: experiments}

To illustrate the derived stability bounds in a practical setting, we consider two test cases: (i) a synthetic setting based on a SBM graph with smooth signals; and (ii) a real-world example using fMRI data \cite{HCP}, where brain connectivity defines the underlying graph structure. In both cases, given the respective graphs the goal is to generate signals that follow the data distribution. For the SBM case, we introduce a controlled synthetic perturbation governed by a parameter $\varepsilon$. For the fMRI case, perturbations arise naturally from errors in estimating the graph structure, specifically from the empirical correlation matrix used to construct the functional connectivity graph. The main objective of this numerical evaluation is to illustrate how loose the derived stability bounds can be in practice and to show that, despite this looseness, the proposed bound-informed regularization leads to more stable models in the presence of graph perturbations.

\begin{figure*}[!t]
\centering
{\includegraphics[width=2.35in]{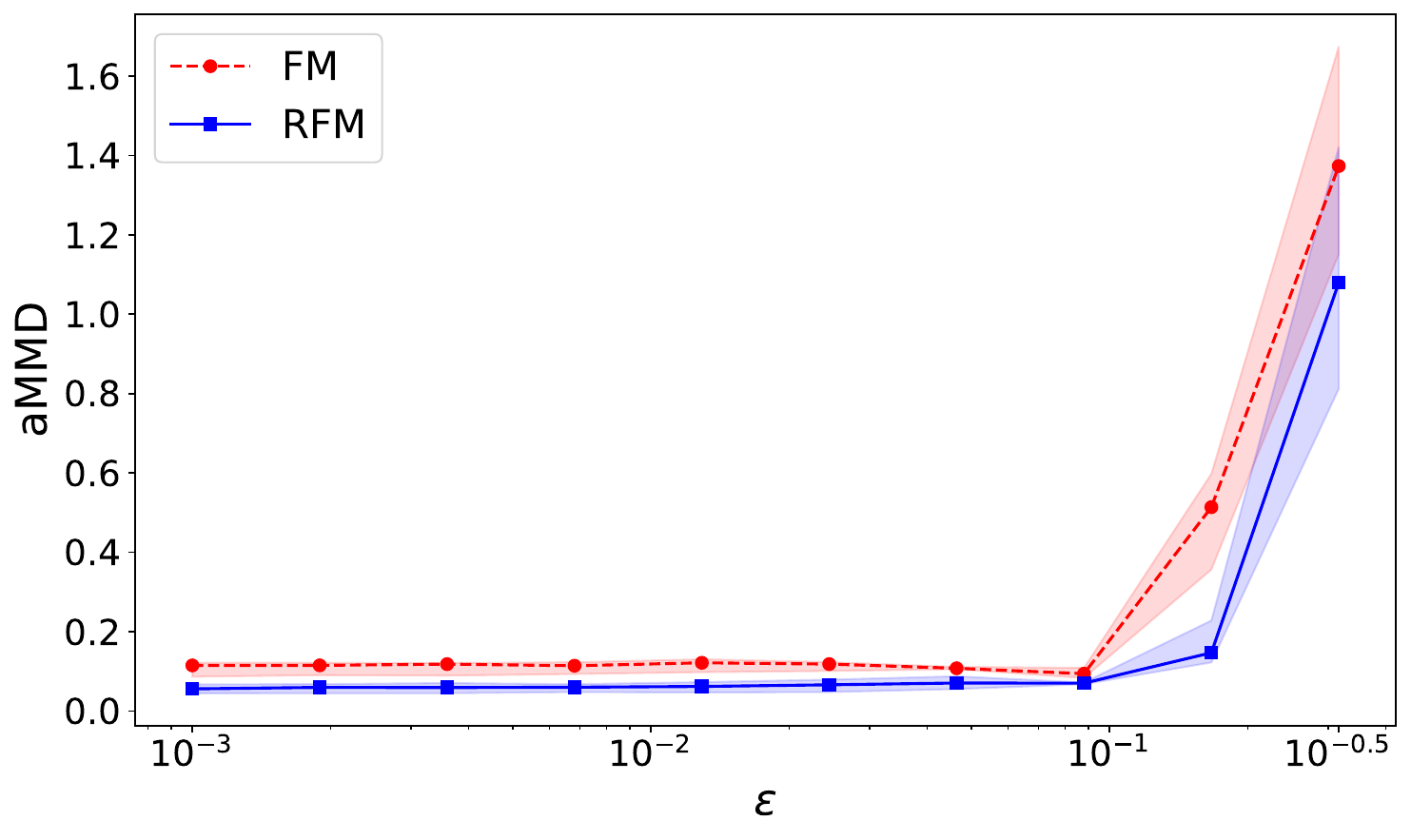}\label{fig:img1}}
\hfil
{\includegraphics[width=2.35in]{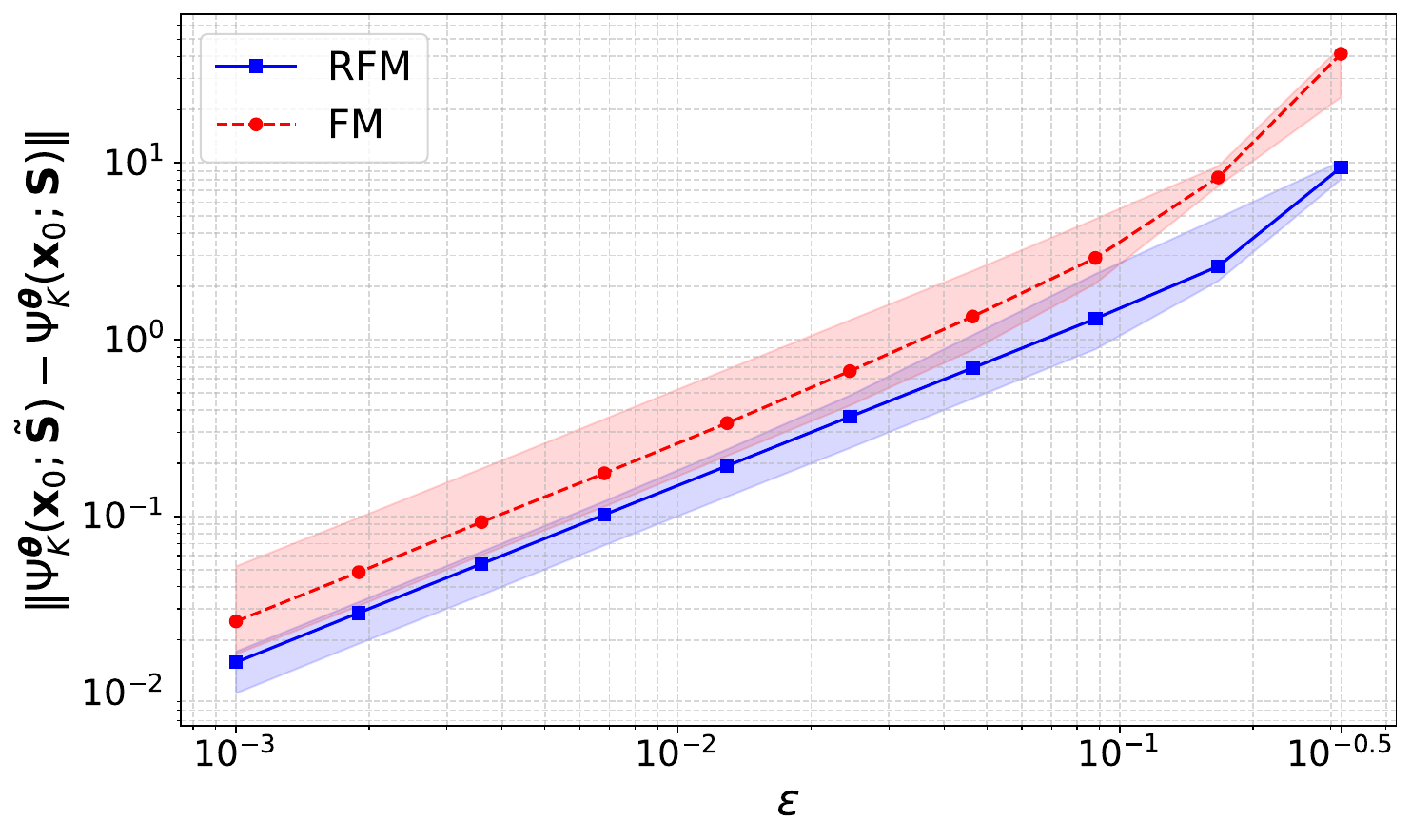}\label{fig:img2}}
\hfil
{\includegraphics[width=2.35in]{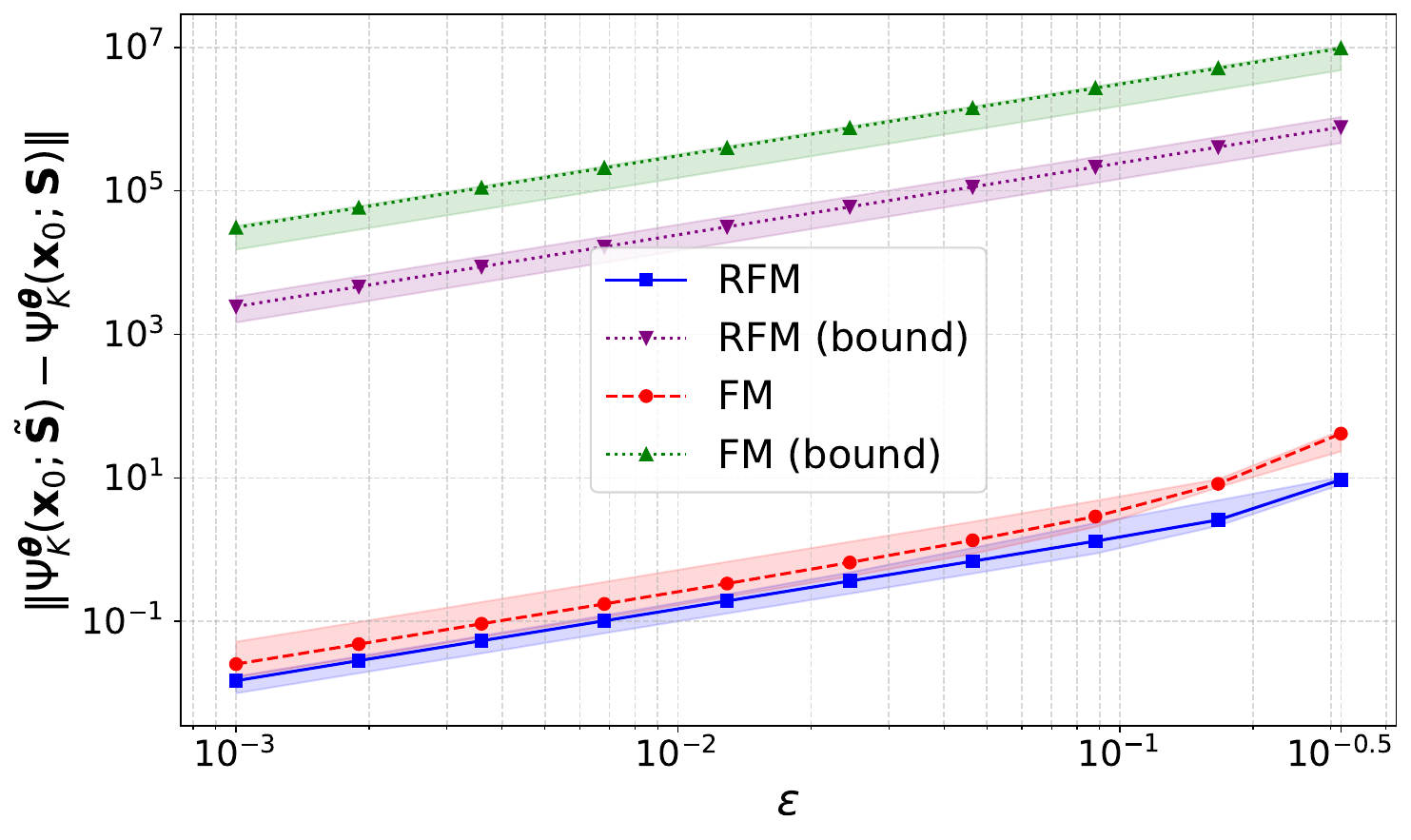}\label{fig:img3}}
\caption{Performance and stability on a SBM graph with smooth signals under a controlled synthetic perturbation of magnitude $\varepsilon$. The aMMD (left) shows that Regularized Flow Matching (RFM) consistently outperforms standard Flow Matching (FM) across all $\varepsilon$ values. The empirical stability (center), measured by the change in the generated output signals due to $\varepsilon$, demonstrates that RFM is more stable than FM. The theoretical stability bounds (right) correctly bound the empirical errors but are not tight.}
\label{fig: sbm}
\end{figure*}

\subsection{Experimental Setup}

\noindent\textbf{Datasets.} In the synthetic setting, we use an SBM graph \cite{holland1983stochastic} with two communities of 10 nodes each ($N=20$). Signals in each community are drawn from a Gaussian distribution with standard deviation 1. One community has mean 1, while the other has mean $-1$. We generate 500 training, 500 validation, and 500 test graph signals. To induce smoothness, the signals are processed using a low-pass graph filter. For the fMRI data, we consider a single subject from the HCP dataset, with data in $\mathbb{R}^{360 \times 1190}$, where 360 represents brain regions (nodes) and 1190 is the number of time points. We treat each time point as an individual graph signal, i.e., a vector $\mathbf{x}^{(i)} \in \mathbb{R}^{360}$. Thus, the dataset consists of $\{\mathbf{x}^{(1)}, \dots, \mathbf{x}^{(1190)}\}$. This provides a well-motivated practical scenario, as fMRI signals are known to inherently exhibit graph signal smoothness over functional brain connectomes \cite{huang2016graph, preti2019decoupling}. We first split the dataset into 80\% training/validation data and 20\% test data. Subsequently, 20\% of the training/validation set is held out for validation, while the remaining 80\% is used for training.\vspace{2pt}

\noindent\textbf{Architecture.} We parameterize the continuous vector field $u_t^{\boldsymbol{\theta}}(\mathbf{x}_t; \mathbf{S})$ defined in \eqref{eq: VF} using a two-layer GNN ($L=2$). The temporal conditioning function is defined as $g(\mathbf{x}, t) = [\mathbf{x}, \mathbf{1}\textbf{emb}(t)^\top]$, where $\textbf{emb}(t) \in \mathbb{R}^{64}$ is a standard 64-dimensional sinusoidal time embedding. Because the input consists of the 1-dimensional node feature and the 64-dimensional time embedding, the input feature dimension is $F_0 = 65$. The GNN hidden layer has a width of $F_1 = 4$ features, and the output layer $F_2 = 1$. Both graph convolutional layers utilize $P=4$ filter taps and use the normalized graph Laplacian as the GSO. We employ a normalized SiLU \cite{elfwing2018sigmoid} as the pointwise nonlinearity $\sigma(\cdot)$. While we opt for this relatively simple architecture to balance expressiveness and stability, we have empirically observed consistent results across a range of alternative architectural configurations. We generate samples using an Euler sampler with $K=100$ steps.\vspace{2pt}

\noindent\textbf{Loss functions.} We consider two training alternatives. For the first case, standard Flow Matching (FM), we directly minimize the conditional flow matching loss in \eqref{eq: CFM} using the optimal transport path \cite{lipman2022flow}, defined as $p_t(\mathbf{x}|\mathbf{z}) = \mathcal{N}(\mathbf{x}; t\mathbf{z}, (1-t)^2 \mathbf{I})$. For the second case, Regularized Flow Matching (RFM), we use the new regularized loss proposed in \eqref{eq: RFM}, which combines \eqref{eq: CFM} with the regularization term introduced in Proposition~\ref{prop: lipschitz-bound}, with $\mu = 0.01$ selected via a grid search.\vspace{2pt}

\noindent\textbf{Training and evaluation.} We use an ADAM optimizer with learning rate $0.001$. We train for 500 epochs with early stopping of 50 epochs and a batch size of 64. In all experiments, the model is trained using the unperturbed graph, i.e., $\varepsilon = 0$, and for the fMRI setting the graph is constructed using the full training and validation data. At test time, we evaluate robustness by introducing graph perturbations. In both cases, we run 10 random datasets; for the SBM this corresponds to drawing new samples, while for the fMRI case we generate new train/validation/test partitions. For each dataset draw, we sample 10 perturbed GSOs $\tilde{\mathbf{S}}$. We report two evaluation metrics: (i) a performance-oriented metric based on the average Maximum Mean Discrepancy (aMMD) \cite{gretton2012kernel}; and (ii) the variation in generated outputs induced by perturbations of the graph support, quantified by $\|\Psi_k^{\boldsymbol{\theta}}(\mathbf{x}_0; \tilde{\mathbf{S}}) - \Psi_k^{\boldsymbol{\theta}}(\mathbf{x}_0; \mathbf{S})\|$. We follow the evaluation protocol of prior work for aMMD \cite{rozada2025graphawarediffusionsignalgeneration}. Specifically, we compute MMD between generated and test distributions over three signal descriptors—quadratic variation, spectral centroid, and degree correlation—capturing smoothness, spectral content, and graph alignment, respectively, and average the resulting MMDs. In all resulting plots, the curves represent the median across these independent runs, and the shaded regions indicate the 25th and 75th percentiles.\vspace{2pt}

\noindent\textbf{Synthetic perturbation.} At test time, we evaluate robustness under controlled perturbations of the graph structure in the SBM setting. Specifically, we generate a random diagonal matrix $\mathbf{E}$ such that $\|\mathbf{E}\|_2 \leq \varepsilon$. Following the framework of \cite{Gama_2020}, the entries of $\mathbf{E}$ are drawn uniformly from the interval $[(1-\varepsilon)\varepsilon, \varepsilon]$. The perturbed GSO is then defined as $\tilde{\mathbf{S}} = \mathbf{S} + \mathbf{E}\mathbf{S} + \mathbf{S}\mathbf{E}$. We simulate logarithmically-spaced $\varepsilon$ values between $10^{-3}$ and $10^{-0.5}$.\vspace{2pt}

\noindent\textbf{Functional connectivity perturbation.} At test time, we further evaluate robustness under data-driven perturbations of the graph in the fMRI setting. We construct the GSO $\tilde{\mathbf{S}}$ from an empirical Pearson correlation matrix estimated on a subset of the available data. Specifically, we vary the fraction of samples used to estimate the correlation network from the training and validation sets, ranging from $1\%$ to $90\%$ in logarithmic scale.

\subsection{Results and Discussion}

\noindent\textbf{Synthetic perturbation on SBM graphs.} Fig.~\ref{fig: sbm} summarizes the results for the synthetic SBM experiment. In terms of generative quality (Fig.~\ref{fig: sbm}, left), the RFM model achieves a lower aMMD than standard FM across all noise levels $\varepsilon$. The empirical stability evaluation (Fig.~\ref{fig: sbm}, center) demonstrates that RFM is consistently more stable than FM. By actively penalizing the spatial Lipschitz constant during training, RFM effectively limits the accumulation of structural errors along the trajectory. When comparing the empirical output variations against the theoretical bounds (Fig.~\ref{fig: sbm}, right), we observe that while the bounds correctly limit the maximum error, they are not tight. This looseness is expected and stems from two primary sources. First, computing the exact Lipschitz constant of a neural network is NP-hard \cite{scaman2019lipschitzregularitydeepneural}, so we rely on the upper bound established in Proposition~\ref{prop: lipschitz-bound}. This approximation is inherently loose because it is a conservative upper bound, and because it bounds the global Lipschitz constant $M$ instead of the tighter one-sided Lipschitz constant $m$. Substituting this worst-case approximation into Theorem~\ref{thm: flow-stability} yields a pessimistic estimate for the exponential growth factor $\Omega_1$.	
Second, the stability bound of the base GNN in \eqref{eq: gnn-stability}, as established in \cite{Gama_2020}, is already a conservative upper bound. This native lack of tightness naturally propagates through our flow bounds.\vspace{2pt}

\begin{figure*}[!t]
\centering
{\includegraphics[width=2.35in]{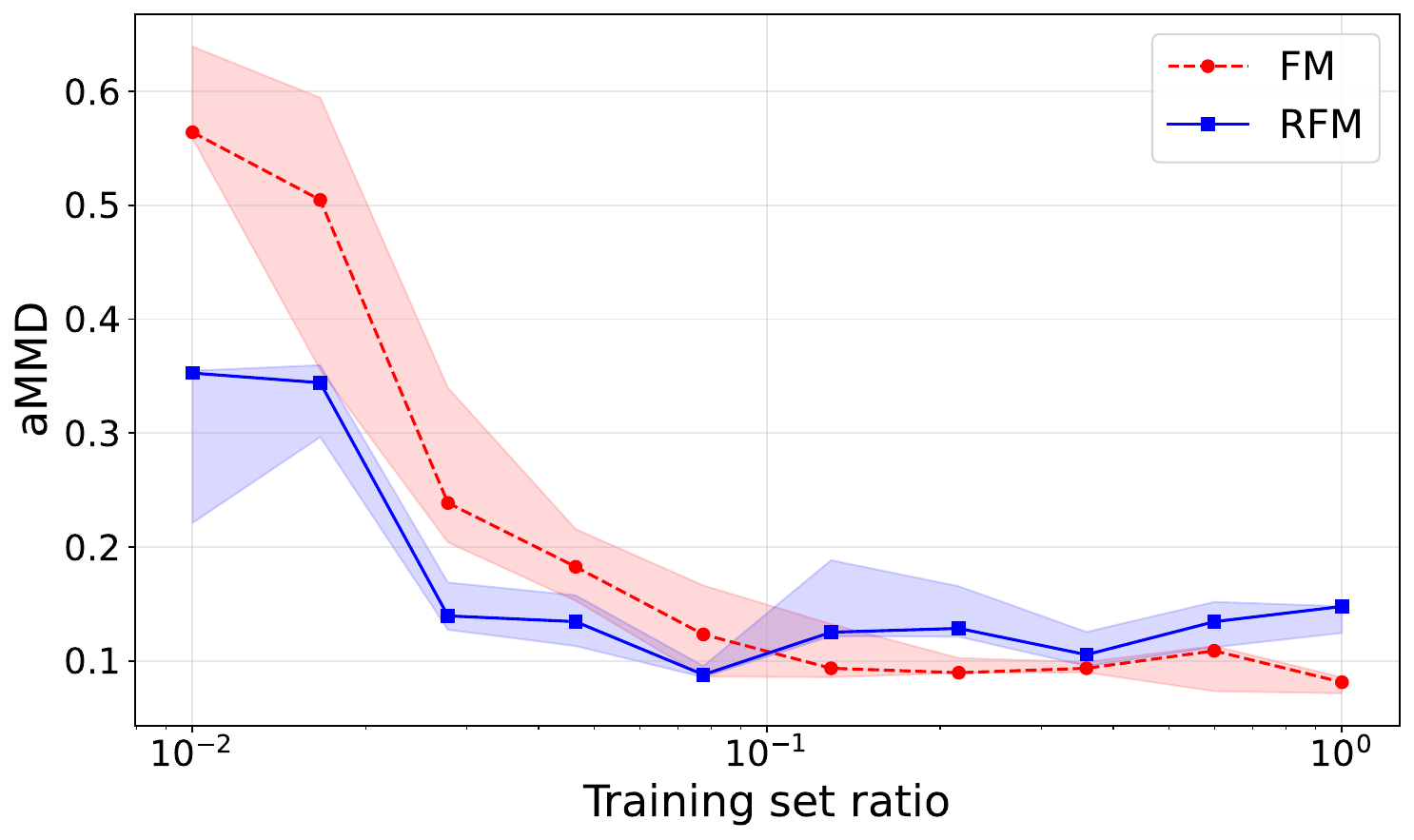}\label{fig:img1}}
\hfil
{\includegraphics[width=2.35in]{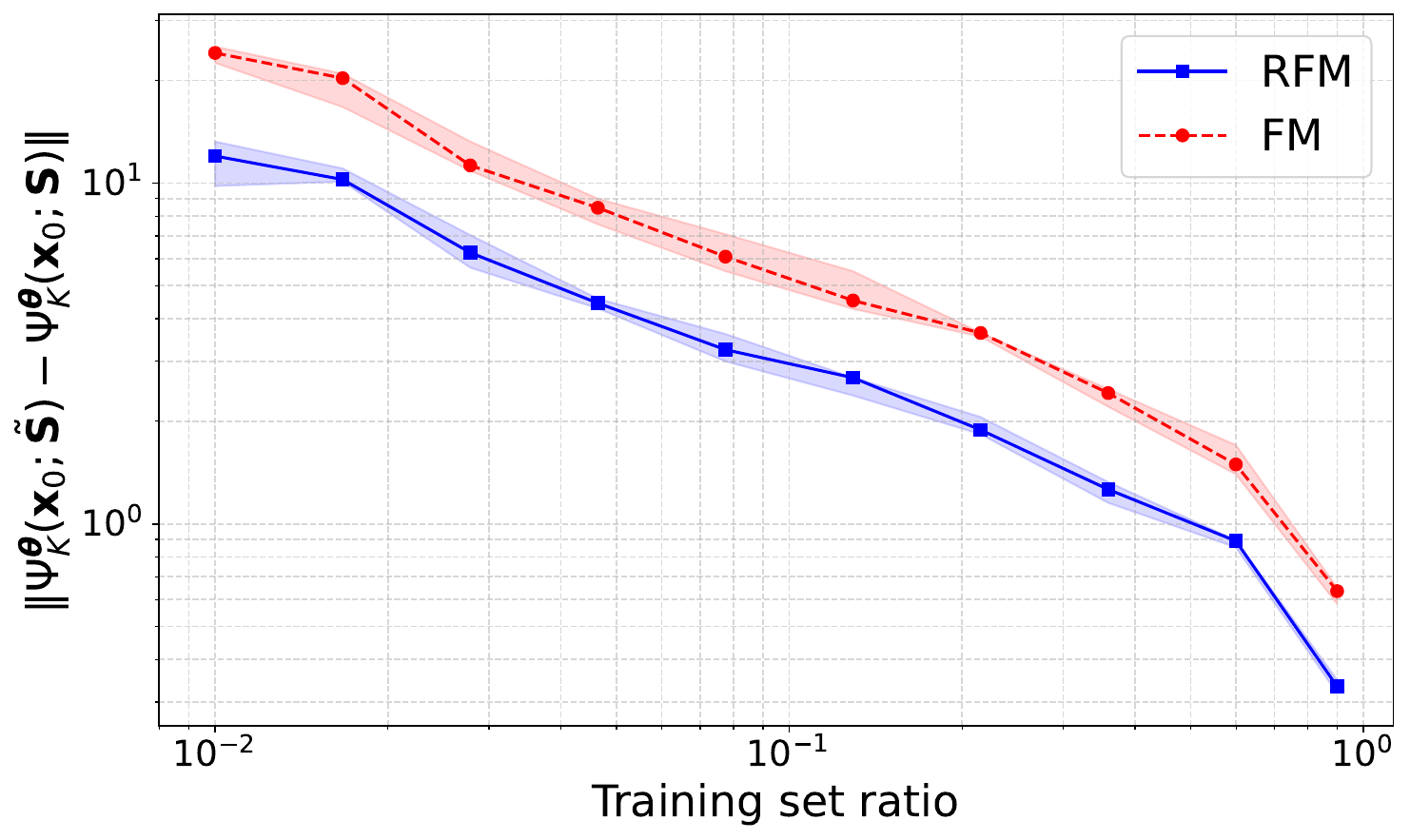}\label{fig:img2}}
\hfil
{\includegraphics[width=2.35in]{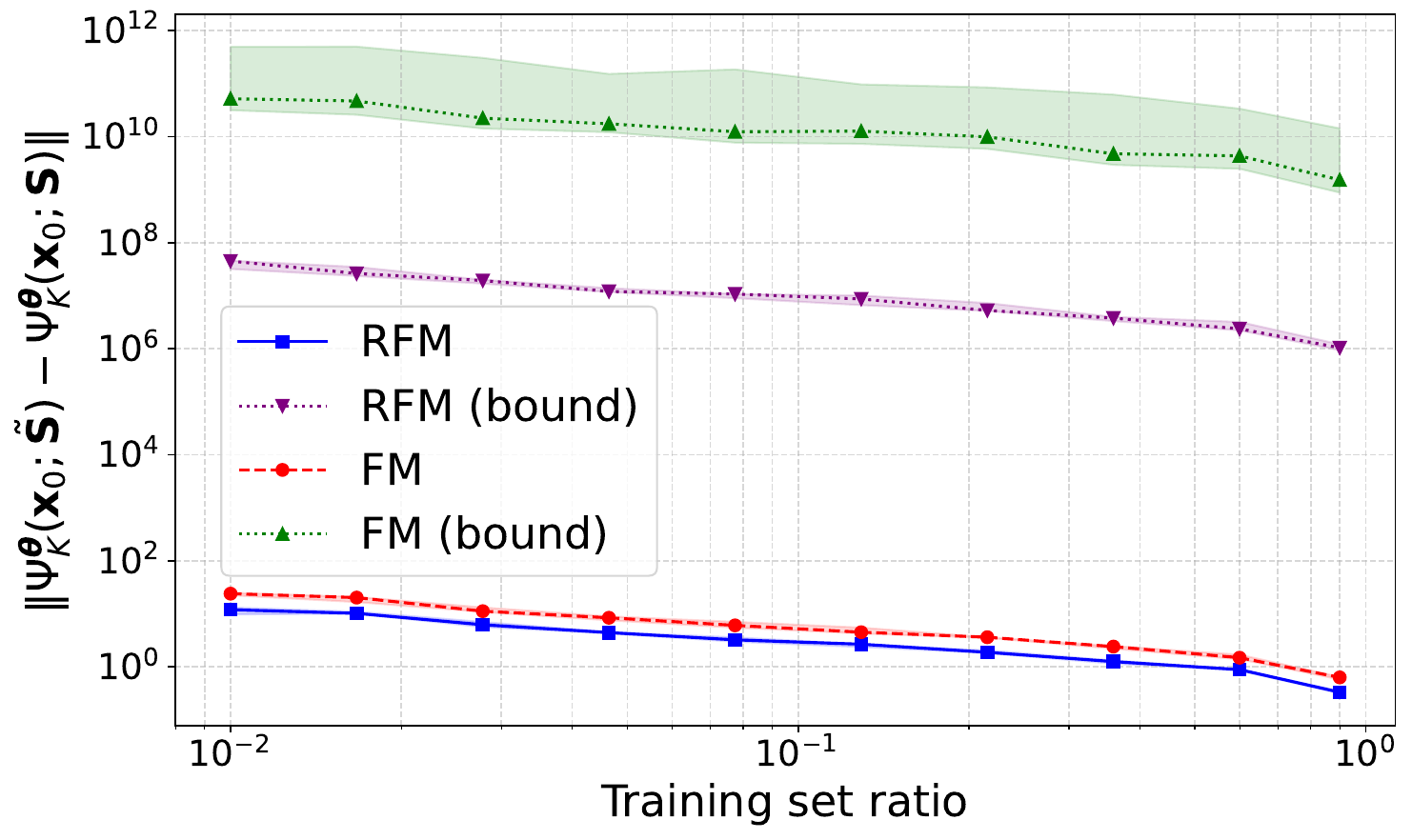}\label{fig:img3}}
\caption{Performance and stability on fMRI signals over a functional brain connectome under data-driven graph perturbations. The x-axis represents the ratio of training samples used to estimate the empirical correlation matrix. The aMMD (left) shows that Regularized Flow Matching (RFM) performs comparably to standard Flow Matching (FM) at high sample ratios, but achieves better performance at lower ratios. The empirical stability (center), measured by the change in the generated output signals due to graph estimation errors, demonstrates that RFM is more stable than FM. The theoretical stability bounds (right) correctly bound the empirical errors but are not tight.}
\label{fig: fmri}
\end{figure*}

\noindent\textbf{Realistic perturbation on fMRI data.} Fig.~\ref{fig: fmri} depicts the fMRI experiments, reporting how these models behave under real-world, data-driven perturbations. As the ratio of training samples decreases, the empirical correlation matrix becomes a noisier estimate of the brain connectome used to train the model. For high sample ratios (Fig.~\ref{fig: fmri}, left), RFM and FM yield comparable aMMD, indicating that the regularization does not degrade performance when the graph structure is reliable. For low sample ratios, RFM achieves better aMMD. The empirical divergence (Fig.~\ref{fig: fmri}, center) shows that RFM is consistently more stable than FM when the graph estimate is degraded. The bounds for the fMRI test case (Fig.~\ref{fig: fmri}, right) exhibit the same conservative behavior discussed previously, but they are notably looser than in the SBM case. This increased looseness is driven by two factors: the brain connectome has a much higher number of nodes, and the complexity of the fMRI signals requires the learned models (both FM and RFM) to have a larger spatial Lipschitz constant $M$. Nevertheless, the same overall general conclusions hold. Crucially, despite the looseness of the theoretical bounds, explicitly controlling the architectural constants derived from them yields a model that is more robust to graph perturbations.

\section{Conclusion}
\label{sec: conclusion}
In this paper, we established the equivariance and stability properties relevant to the generation of graph signals using GNN-parametrized CNFs. We proved that both the continuous-time ODEs and their discrete numerical approximations, specifically Euler and Heun samplers, preserve permutation equivariance inherited from the underlying GNN. By analyzing how structural errors propagate through the generative flow, we derived explicit Wasserstein stability bounds for the generated distributions under relative graph perturbations.

Beyond theoretical guarantees, these bounds informed a practical training strategy. Having identified the spatial Lipschitz constant of the vector field as the main source of error accumulation, we introduced RFM to actively penalize this constant. Our experiments on synthetic SBMs and empirical fMRI data confirmed that explicitly controlling this constant yields generative models that are more robust to structural noise, without sacrificing the quality of the generated signals. While our practical implementation focused specifically on regularizing the spatial Lipschitz constant, exploring the theoretical and empirical compromises of jointly regularizing the spatial and frequency-response Lipschitz constants represents one direction to further optimize the trade-off between model expressiveness and structural robustness.

The analytical framework presented here establishes a foundation for evaluating broader classes of continuous generative models. Because the continuous-time and discrete stability guarantees are grounded in the properties of the parameterized vector field, this analysis can be extended to stochastic differential equations, provided the injected noise satisfies permutation equivariance. Furthermore, our results provide the basis to establish transferability bounds, addressing the problem of training a generative model on smaller (sampled) graphs and transferring them to large-scale graphs \cite{juan2023tsp}.
\appendices

%
%

\section{Permutation Equivariance of the Continuous Flow}\label{proof: flow-equivariance}

\begin{proof}[Proof of Proposition \ref{prop: flow-equivariance}]
Let $\mathbf{x}_t = \Phi_t^{\boldsymbol{\theta}}(\mathbf{x}_0; \mathbf{S})$ and define $\mathbf{y}_t \coloneqq \mathbf{P}^\top \mathbf{x}_t$. Differentiating with respect to $t$ and using the equivariance of the vector field $u_t^{\boldsymbol{\theta}}$ in \eqref{eq: VF-equivariance} yields
\begin{equation*}
    \begin{aligned}
    \frac{d}{dt}\mathbf{y}_t &= \mathbf{P}^\top \frac{d}{dt}\mathbf{x}_t = \mathbf{P}^\top u_t^{\boldsymbol{\theta}}(\mathbf{x}_t; \mathbf{S}) \\
    &= u_t^{\boldsymbol{\theta}}(\mathbf{P}^\top \mathbf{x}_t; \mathbf{P}^\top \mathbf{S} \mathbf{P}) = u_t^{\boldsymbol{\theta}}(\mathbf{y}_t; \hat{\mathbf{S}}).
\end{aligned}
\end{equation*}
Since $\mathbf{y}_0 = \mathbf{P}^\top \mathbf{x}_0 = \hat{\mathbf{x}}_0$, the path $\mathbf{y}_t$ satisfies the initial value problem defining $\Phi_t^{\boldsymbol{\theta}}(\hat{\mathbf{x}}_0; \hat{\mathbf{S}})$. By the uniqueness of solutions guaranteed by the Picard-Lindelöf theorem, it follows that $\mathbf{y}_t = \Phi_t^{\boldsymbol{\theta}}(\hat{\mathbf{x}}_0; \hat{\mathbf{S}})$, which concludes the proof.
\end{proof}

%
%

\section{Stability of the Flow Model}
\label{proof: flow-stability}

\begin{proof}[Proof of Theorem \ref{thm: flow-stability}]
Let $\mathbf{P}_0 \in \mathcal{P}$ be the permutation matrix from Section \ref{sec: stability-of-gnn} that defines the relative perturbation in \eqref{eq: perturbation}. Let $\tilde{\mathbf{x}}_t \coloneqq \Phi_t^{\boldsymbol{\theta}}(\mathbf{x}_0; \tilde{\mathbf{S}})$ and $\mathbf{x}_t \coloneqq \Phi_t^{\boldsymbol{\theta}}(\mathbf{P}_0^\top \mathbf{x}_0; \mathbf{S})$.

First, we upper bound the minimum over all permutations by evaluating the error specifically at $\mathbf{P}_0$, to obtain
\begin{equation*}
    \begin{aligned}
        \min_{\mathbf{P} \in \mathcal{P}} &\|\mathbf{P}^\top \Phi_t^{\boldsymbol{\theta}}( \mathbf{x}_0; \tilde{\mathbf{S}})- \Phi_t^{\boldsymbol{\theta}}(\mathbf{P}^\top \mathbf{x}_0; \mathbf{S}) \| \\
        &\leq \|\mathbf{P}_0^\top \Phi_t^{\boldsymbol{\theta}}( \mathbf{x}_0; \tilde{\mathbf{S}})- \Phi_t^{\boldsymbol{\theta}}(\mathbf{P}_0^\top \mathbf{x}_0; \mathbf{S}) \|\\
        & = \|\mathbf{P}_0^\top \tilde{\mathbf{x}}_t - \mathbf{x}_t\|.
    \end{aligned}
\end{equation*}

Let $\mathbf{z}_t \coloneqq \mathbf{P}_0^\top \tilde{\mathbf{x}}_t - \mathbf{x}_t$. Differentiating yields
\begin{equation*}
    \begin{aligned}
        \frac{1}{2}\frac{d}{dt}\|\mathbf{z}_t\|^2 
        &= \frac{1}{2} \frac{d}{dt} \langle \mathbf{z}_t, \mathbf{z}_t \rangle = \langle \mathbf{z}_t, \frac{d}{dt} \mathbf{z}_t \rangle  \\
        &= \langle \mathbf{P}_0^\top \tilde{\mathbf{x}}_t - \mathbf{x}_t , \mathbf{P}_0^\top u_t^{\boldsymbol{\theta}} (\tilde{\mathbf{x}}_t; \tilde{\mathbf{S}}) - u_t^{\boldsymbol{\theta}} (\mathbf{x}_t;\mathbf{S})\rangle.
    \end{aligned}
\end{equation*}
By adding and subtracting the term $\mathbf{P}_0^\top u_t^{\boldsymbol{\theta}}(\mathbf{P}_0\mathbf{x}_t; \tilde{\mathbf{S}})$, we can split the inner product
\begin{equation*}
    \begin{aligned}
        \frac{1}{2}\frac{d}{dt}\|\mathbf{z}_t\|^2 
        &= \langle \mathbf{P}_0^\top \tilde{\mathbf{x}}_t - \mathbf{x}_t, \mathbf{P}_0^\top\big( u_t^{\boldsymbol{\theta}}(\tilde{\mathbf{x}}_t; \tilde{\mathbf{S}}) - u_t^{\boldsymbol{\theta}} (\mathbf{P}_0\mathbf{x}_t; \tilde{\mathbf{S}}) \big) \\
        &+ \mathbf{P}_0^\top u_t^{\boldsymbol{\theta}}(\mathbf{P}_0 \mathbf{x}_t; \tilde{\mathbf{S}}) - u_t^{\boldsymbol{\theta}}(\mathbf{x}_t;\mathbf{S})\rangle \\
        &= \langle \mathbf{P}_0^\top \tilde{\mathbf{x}}_t - \mathbf{x}_t , \mathbf{P}_0^\top\big( u_t^{\boldsymbol{\theta}}(\tilde{\mathbf{x}}_t; \tilde{\mathbf{S}}) - u_t^{\boldsymbol{\theta}}(\mathbf{P}_0\mathbf{x}_t; \tilde{\mathbf{S}})\big)\rangle \\
        & \quad + \langle \mathbf{P}_0^\top \tilde{\mathbf{x}}_t - \mathbf{x}_t , \mathbf{P}_0^\top u_t^{\boldsymbol{\theta}}(\mathbf{P}_0\mathbf{x}_t; \tilde{\mathbf{S}})- u_t^{\boldsymbol{\theta}}(\mathbf{x}_t; \mathbf{S})\rangle.
    \end{aligned}
\end{equation*}

To bound this expression, we handle each term separately. For the first term, we use the fact that the permutation matrix $\mathbf{P}_0$ is orthogonal (i.e., $\langle \mathbf{P}_0 \mathbf{a}, \mathbf{P}_0 \mathbf{b} \rangle = \langle \mathbf{a}, \mathbf{b} \rangle$) and apply the one-sided Lipschitz condition (\ref{eq: one-sided lipschitz}). For the second term, we apply the Cauchy-Schwarz inequality and we use the stability of the vector field by substituting the input signal with $\mathbf{P}_0\mathbf{x}_t$ in (\ref{eq: VF-stability}). This yields
\begin{equation*}
    \begin{aligned}
        \frac{1}{2}\frac{d}{dt}\|\mathbf{z}_t\|^2 
        &= \langle \tilde{\mathbf{x}}_t - \mathbf{P}_0 \mathbf{x}_t , u_t^{\boldsymbol{\theta}}(\tilde{\mathbf{x}}_t; \tilde{\mathbf{S}}) - u_t^{\boldsymbol{\theta}}(\mathbf{P}_0\mathbf{x}_t; \tilde{\mathbf{S}})\rangle \\ 
        & + \langle \mathbf{z}_t , \mathbf{P}_0^\top u_t^{\boldsymbol{\theta}}(\mathbf{P}_0\mathbf{x}_t; \tilde{\mathbf{S}})- u_t^{\boldsymbol{\theta}}(\mathbf{x}_t; \mathbf{S})\rangle \\
        &\leq m_t \|\tilde{\mathbf{x}}_t - \mathbf{P}_0 \mathbf{x}_t\|^2 \\
        & + \|\mathbf{z}_t\| \big(\Gamma \varepsilon + \mathcal{O}(\varepsilon^2)\big) \|g(\mathbf{x}_t,t)\|.
    \end{aligned}
\end{equation*}
Since $\|\tilde{\mathbf{x}}_t - \mathbf{P}_0 \mathbf{x}_t\| = \|\mathbf{P}_0^\top \tilde{\mathbf{x}}_t - \mathbf{x}_t\| = \|\mathbf{z}_t\|$, the inequality simplifies to
\begin{align*}
    \frac{1}{2}\frac{d}{dt}\|\mathbf{z}_t\|^2 \leq m_t \|\mathbf{z}_t\|^2 + \|\mathbf{z}_t\|\big(\Gamma \varepsilon + \mathcal{O}(\varepsilon^2)\big) \|g(\mathbf{x}_t,t)\|.
\end{align*}

By the definition of $C_g(\mathbf{x}_0)$, which takes the supremum over the time interval and the maximum over all permutation matrices in $\mathcal{P}$, it directly follows that $\|g(\mathbf{x}_t,t) \| \leq C_g(\mathbf{x}_0)$, and hence 
\begin{align*}
    \frac{1}{2}\frac{d}{dt}\|\mathbf{z}_t\|^2  \leq m_t \|\mathbf{z}_t\|^2 + \|\mathbf{z}_t\|\big(\Gamma \varepsilon + \mathcal{O}(\varepsilon^2)\big) C_g(\mathbf{x}_0).
\end{align*}
Let $r_t = \|\mathbf{z}_t\|$. From the chain rule $\frac{1}{2}\frac{d}{dt}r_t^2 = r_t \frac{d}{dt}r_t$, and we thus obtain
\begin{align*}
    r_t \frac{d}{dt}r_t \leq m_t r_t^2 + r_t\big(\Gamma \varepsilon + \mathcal{O}(\varepsilon^2)\big) C_g(\mathbf{x}_0).
\end{align*}
For $t$ such that $r_t = 0$, the result is trivially true. For $r_t>0$, we can divide by $r_t$ to obtain
\begin{align*}
    \frac{d}{dt}r_t \leq m_t r_t + \big(\Gamma \varepsilon + \mathcal{O}(\varepsilon^2)\big) C_g(\mathbf{x}_0).
\end{align*}
Using Grönwall's inequality and the fact that the initial error $r_0 = \|\mathbf{P}_0^\top \mathbf{x}_0 - \mathbf{P}_0^\top \mathbf{x}_0\| = 0$, we have
\begin{align*}
    r_t \leq C_g(\mathbf{x}_0) \big(\Gamma \varepsilon+ \mathcal{O}(\varepsilon^2)\big)\left( \int_0^t \operatorname{exp}\left(\int_\tau^t m_s ds \right)d\tau \right), 
\end{align*}
which concludes the proof.
\end{proof}

%
%

\section{Wasserstein Stability of the Flow Model}
\label{proof: wasserstein-stability}

\begin{proof}[Proof of Corollary \ref{cor: wasserstein-stability}]
Let $\mathbf{P}_0 \in \mathcal{P}$ be the specific permutation matrix from Section \ref{sec: stability-of-gnn} that minimizes the relative perturbation error in \eqref{eq: perturbation}. Evaluating the flow on the exact same initial random variable $\mathbf{x}_0$ under both the nominal graph $\mathbf{S}$ and the perturbed graph $\tilde{\mathbf{S}}$ forms a valid coupling. Thus, we can upper bound the Wasserstein-2 distance using the bound derived in Theorem \ref{thm: flow-stability}
\begin{equation*}
    \begin{aligned}
        \min_{\mathbf{P} \in \mathcal{P}}W_2( \tilde{\pi}_t^{\mathbf{P}}, \pi_t^{\mathbf{P}} ) ^2 
        &\leq \mathbb{E}_{\mathbf{x}_0}\left[ \|\mathbf{P}^\top_0 \Phi_t^{\boldsymbol{\theta}}(\mathbf{x}_0;\tilde{\mathbf{S}}) - \Phi_t^{\boldsymbol{\theta}}(\mathbf{P}^\top_0 \mathbf{x}_0;\mathbf{S})\|^2 \right] \\
        &\leq \mathbb{E}_{\mathbf{x}_0}\left[ \Big(\Omega_t C_g(\mathbf{x}_0) \big(\Gamma \varepsilon + \mathcal{O}(\varepsilon^2)\big)\Big)^2 \right].
    \end{aligned}
\end{equation*}
Because $\Omega_t$ and $\big(\Gamma \varepsilon + \mathcal{O}(\varepsilon^2)\big)$ are deterministic and independent of $\mathbf{x}_0$, we can factor them out of the expectation
\begin{equation*}
    \begin{aligned}
        \mathbb{E}_{\mathbf{x}_0} &\left[ \Big(\Omega_t C_g(\mathbf{x}_0) \big(\Gamma \varepsilon + \mathcal{O}(\varepsilon^2)\big)\Big)^2 \right] \\
        &= \Omega_t^2 \big(\Gamma \varepsilon + \mathcal{O}(\varepsilon^2)\big)^2 \mathbb{E}_{\mathbf{x}_0}\left[ C_g(\mathbf{x}_0)^2 \right] \\
        &= \Omega_t^2 \big(\Gamma \varepsilon + \mathcal{O}(\varepsilon^2)\big)^2 (C_g')^2.
    \end{aligned}
\end{equation*}
Taking the square root of both sides directly yields \eqref{eq:bound_cor_2}.
%
\end{proof}

%
%

\section{Permutation Equivariance of the Discrete Flow}
\label{proof: discrete-flow-equivariance}

\begin{proof}[Proof of Proposition \ref{prop: discrete-flow-equivariance}]
We proceed by induction on $k$. For the base case $k=0$, the definition of the discrete flow in \eqref{eq: discrete-flow-map} and our initial condition yields
\begin{align*}
    \Psi_0^{\boldsymbol{\theta}}(\hat{\mathbf{x}}_0; \hat{\mathbf{S}}) = \hat{\mathbf{x}}_0 = \mathbf{P}^\top \mathbf{x}_0 = \mathbf{P}^\top \Psi_0^{\boldsymbol{\theta}}(\mathbf{x}_0; \mathbf{S}).
\end{align*}

Assume the claim holds for some $k \in \{0, \dots, K-1\}$, meaning $\Psi_k^{\boldsymbol{\theta}}(\hat{\mathbf{x}}_0; \hat{\mathbf{S}}) = \mathbf{P}^\top \Psi_k^{\boldsymbol{\theta}}(\mathbf{x}_0; \mathbf{S})$. For step $k+1$, we apply the recursive definition of the discrete flow and the permutation equivariance of the update map $T_k$ in \eqref{eq: sampler-equivariance} to obtain
\begin{equation*}
    \begin{aligned}
        \Psi_{k+1}^{\boldsymbol{\theta}}(\hat{\mathbf{x}}_0; \hat{\mathbf{S}}) &= T_k^{\mathbf\theta}\big(\Psi_k^{\boldsymbol{\theta}}(\hat{\mathbf{x}}_0; \hat{\mathbf{S}}); \hat{\mathbf{S}}\big) \\
        &= T_k^{\boldsymbol{\theta}}\big(\mathbf{P}^\top \Psi_k^{\boldsymbol{\theta}}(\mathbf{x}_0; \mathbf{S}); \hat{\mathbf{S}}\big) \\
        &= \mathbf{P}^\top T_k^{\boldsymbol{\theta}}\big(\Psi_k^{\boldsymbol{\theta}}(\mathbf{x}_0; \mathbf{S}); \mathbf{S}\big) \\
        &= \mathbf{P}^\top \Psi_{k+1}^{\boldsymbol{\theta}}(\mathbf{x}_0; \mathbf{S}).
    \end{aligned}
\end{equation*}
By induction, the property holds for all $k \in \{0, \dots, K\}$.
\end{proof}

%
%

\section{Permutation Equivariance of the Euler Sampler}
\label{proof: euler-equivariance}

\begin{proof}
Assuming the vector field $u_k^{\boldsymbol{\theta}}$ is permutation equivariant, we evaluate the Euler update map \eqref{eq: euler-update} on the permuted state $\hat{\mathbf{x}} = \mathbf{P}^\top \mathbf{x}$ and permuted graph $\hat{\mathbf{S}} = \mathbf{P}^\top \mathbf{S} \mathbf{P}$. This yields
\begin{equation*}
    \begin{aligned}
        T_k^{\boldsymbol{\theta}}(\hat{\mathbf{x}}; \hat{\mathbf{S}}) 
        &= \mathbf{P}^\top \mathbf{x} + h\,u^{\boldsymbol{\theta}}_{k}(\mathbf{P}^\top \mathbf{x}; \mathbf{P}^\top \mathbf{S} \mathbf{P}) \\
        &= \mathbf{P}^\top \mathbf{x} + h\,\mathbf{P}^\top u^{\boldsymbol{\theta}}_{k}(\mathbf{x}; \mathbf{S}) \\
        &= \mathbf{P}^\top \big(\mathbf{x} + h\,u^{\boldsymbol{\theta}}_{k}(\mathbf{x};\mathbf{S})\big) \\
        &= \mathbf{P}^\top T_k^{\boldsymbol{\theta}}(\mathbf{x};\mathbf{S}).
    \end{aligned}
\end{equation*}
So, the Euler sampler is permutation equivariant.
\end{proof}

%
%

\section{Permutation Equivariance of the Heun Sampler}
\label{proof: heun-equivariance}

\begin{proof}
Assuming the vector field $u_k^{\boldsymbol{\theta}}$ is permutation equivariant, we evaluate the Heun update map \eqref{eq: heun-update} on the permuted inputs $\hat{\mathbf{x}} = \mathbf{P}^\top \mathbf{x}$ and $\hat{\mathbf{S}} = \mathbf{P}^\top \mathbf{S} \mathbf{P}$. Using the equivariance property on both evaluations of the vector field yields
\begin{equation*}
    \begin{aligned}
        &T_k^{\boldsymbol{\theta}}(\hat{\mathbf{x}}; \hat{\mathbf{S}}) 
        = \hat{\mathbf{x}} + \frac{h}{2} \Big( u_k^{\boldsymbol{\theta}} (\hat{\mathbf{x}};\hat{\mathbf{S}}) + u_{k+1}^{\boldsymbol{\theta}}\big(\hat{\mathbf{x}}+hu_k^{\boldsymbol{\theta}}(\hat{\mathbf{x}};\hat{\mathbf{S}});\hat{\mathbf{S}}\big) \Big) \\
        &= \mathbf{P}^\top \mathbf{x} + \frac{h}{2} \Big( \mathbf{P}^\top u_k^{\boldsymbol{\theta}}(\mathbf{x};\mathbf{S}) + u_{k+1}^{\boldsymbol{\theta}} \big( \mathbf{P}^\top(\mathbf{x}+hu_k^{\boldsymbol{\theta}}(\mathbf{x};\mathbf{S})); \hat{\mathbf{S}} \big) \Big)\\
        &= \mathbf{P}^\top \mathbf{x} + \frac{h}{2} \Big( \mathbf{P}^\top u_k^{\boldsymbol{\theta}}(\mathbf{x};\mathbf{S}) + \mathbf{P}^\top u_{k+1}^{\boldsymbol{\theta}}\big(\mathbf{x}+hu_k^{\boldsymbol{\theta}}(\mathbf{x};\mathbf{S}); \mathbf{S}\big) \Big)\\
        &= \mathbf{P}^\top \left( \mathbf{x} + \frac{h}{2} \Big( u_k^{\boldsymbol{\theta}}(\mathbf{x};\mathbf{S}) + u_{k+1}^{\boldsymbol{\theta}}\big(\mathbf{x}+hu_k^{\boldsymbol{\theta}}(\mathbf{x};\mathbf{S}); \mathbf{S}\big) \Big) \right)\\
        &= \mathbf{P}^\top T_k^{\boldsymbol{\theta}}(\mathbf{x};\mathbf{S}).
    \end{aligned}
\end{equation*}
Thus, the Heun sampler is permutation equivariant as well.
\end{proof}

%
%

\section{Stability of the Discrete Flow}
\label{proof: discrete-flow-stability}

\begin{proof}[Proof of Theorem \ref{thm: discrete-flow-stability}]
Let $\mathbf{P}_0 \in \mathcal{P}$ be the specific permutation matrix from Section \ref{sec: stability-of-gnn} that minimizes the relative perturbation error in \eqref{eq: perturbation}. Define $\tilde{\mathbf{x}}_k \coloneqq \Psi_k^{\boldsymbol{\theta}}(\mathbf{x}_0; \tilde{\mathbf{S}})$ and $\mathbf{x}_k \coloneqq \Psi_k^{\boldsymbol{\theta}}(\mathbf{P}_0^\top \mathbf{x}_0; \mathbf{S})$. By definition of the discrete flow
\begin{equation*}
    \begin{aligned}
        \min_{\mathbf{P} \in \mathcal{P}} &\|\mathbf{P}^\top \Psi_{k+1}^{\boldsymbol{\theta}}(\mathbf{x}_0; \tilde{\mathbf{S}}) - \Psi_{k+1}^{\boldsymbol{\theta}}(\mathbf{P}^\top \mathbf{x}_0; \mathbf{S})\| \\
        &\leq \|\mathbf{P}_0^\top \Psi_{k+1}^{\boldsymbol{\theta}}(\mathbf{x}_0;\tilde{\mathbf{S}}) - \Psi_{k+1}^{\boldsymbol{\theta}}(\mathbf{P}_0^\top \mathbf{x}_0;\mathbf{S})\| \\
        &= \|\mathbf{P}_0^\top \tilde{\mathbf{x}}_{k+1} - \mathbf{x}_{k+1}\| \\
        &= \|\mathbf{P}_0^\top T_k^{\boldsymbol{\theta}} (\tilde{\mathbf{x}}_k; \tilde{\mathbf{S}}) - T_k^{\boldsymbol{\theta}} (\mathbf{x}_k; \mathbf{S})\|.
    \end{aligned}
\end{equation*}

By adding and subtracting the term $\mathbf{P}_0^\top T_k^{\boldsymbol{\theta}}(\mathbf{P}_0 \mathbf{x}_k; \tilde{\mathbf{S}})$, applying the triangle inequality, and using the orthogonality of the permutation matrix $\mathbf{P}_0^\top$, we can bound the error as
\begin{equation*}
    \begin{aligned}
        \|\mathbf{P}_0^\top T_k^{\boldsymbol{\theta}} (\tilde{\mathbf{x}}_k; \tilde{\mathbf{S}}) &- T_k^{\boldsymbol{\theta}} (\mathbf{x}_k; \mathbf{S})\| \\
        &\leq \|\mathbf{P}_0^\top T_k^{\boldsymbol{\theta}}(\tilde{\mathbf{x}}_k; \tilde{\mathbf{S}}) - \mathbf{P}_0^\top T_k^{\boldsymbol{\theta}}(\mathbf{P}_0 \mathbf{x}_k; \tilde{\mathbf{S}}) \| \\
        &\quad + \|\mathbf{P}_0^\top T_k^{\boldsymbol{\theta}}(\mathbf{P}_0 \mathbf{x}_k; \tilde{\mathbf{S}}) - T_k^{\boldsymbol{\theta}} (\mathbf{x}_k; \mathbf{S})\| \\
        &= \|T_k^{\boldsymbol{\theta}}(\tilde{\mathbf{x}}_k; \tilde{\mathbf{S}}) - T_k^{\boldsymbol{\theta}}(\mathbf{P}_0 \mathbf{x}_k; \tilde{\mathbf{S}}) \| \\
        &\quad + \|\mathbf{P}_0^\top T_k^{\boldsymbol{\theta}}(\mathbf{P}_0 \mathbf{x}_k; \tilde{\mathbf{S}}) - T_k^{\boldsymbol{\theta}} (\mathbf{x}_k; \mathbf{S})\|.
    \end{aligned}
\end{equation*}
Because the sampler is state stable (Definition \ref{def: state-stable}), we have
\begin{equation*}
    \begin{aligned}
    \|T_k^{\boldsymbol{\theta}}(\tilde{\mathbf{x}}_k; \tilde{\mathbf{S}}) - T_k^{\boldsymbol{\theta}}(\mathbf{P}_0 \mathbf{x}_k; \tilde{\mathbf{S}}) \| &\leq \alpha \|\tilde{\mathbf{x}}_k - \mathbf{P}_0\mathbf{x}_k\| \\
    &= \alpha \|\mathbf{P}_0^\top \tilde{\mathbf{x}}_k - \mathbf{x}_k\|.
\end{aligned}
\end{equation*}
By substituting $\mathbf{P}_0\mathbf{x}_k$ into the definition of a graph stable sampler (Definition \ref{def: graph-stable}) and using the permutation equivariance of $\beta$ (i.e., $\beta(\mathbf{P}_0\mathbf{x}_k, t_k) = \beta(\mathbf{x}_k, t_k)$), yields
\begin{align*}
    \|\mathbf{P}_0^\top T_k^{\boldsymbol{\theta}}(\mathbf{P}_0 \mathbf{x}_k; \tilde{\mathbf{S}}) - T_k^{\boldsymbol{\theta}} (\mathbf{x}_k; \mathbf{S})\| \leq \beta(\mathbf{x}_k, t_k) \big(\Gamma\varepsilon + \mathcal{O}(\varepsilon^2)\big).
\end{align*}
By definition, $\mathbf{x}_k = \Psi_k^{\boldsymbol{\theta}}(\mathbf{P}_0^\top \mathbf{x}_0; \mathbf{S})$ is the unperturbed trajectory starting from the permuted initial condition $\mathbf{P}_0^\top \mathbf{x}_0$. Because $\beta_{\max}$ is defined as the maximum over all $\mathbf{P} \in \mathcal{P}$, it directly follows that $\beta(\mathbf{x}_k, t_k) \leq \beta_{\max}$. Thus, the error bound satisfies the recurrence relation
\begin{equation*}
    \begin{aligned}
        \|\mathbf{P}_0^\top \tilde{\mathbf{x}}_{k+1} - \mathbf{x}_{k+1}\| \leq \alpha \|\mathbf{P}_0^\top \tilde{\mathbf{x}}_k - \mathbf{x}_k\| + \beta_{\max} \big(\Gamma\varepsilon + \mathcal{O}(\varepsilon^2)\big).
    \end{aligned}
\end{equation*}
Since $\|\mathbf{P}_0^\top \tilde{\mathbf{x}}_{0} - \mathbf{x}_{0}\| =0$, iterating this recurrence relation yields a geometric series
\begin{equation*}
    \begin{aligned}
    \|\mathbf{P}_0^\top \tilde{\mathbf{x}}_{k} - \mathbf{x}_{k}\| 
    &\leq \beta_{\max} \big(\Gamma\varepsilon + \mathcal{O}(\varepsilon^2)\big) \sum_{i=0}^{k-1} \alpha^i 
    \end{aligned}
\end{equation*}
\begin{equation*}
    \begin{aligned}
    &= \begin{cases}
    \frac{1 - \alpha^k}{1-\alpha}\beta_{\max} \big(\Gamma\varepsilon + \mathcal{O}(\varepsilon^2)\big) & \text{if } \alpha \neq 1, \\[8pt]
    k \beta_{\max} \big(\Gamma\varepsilon + \mathcal{O}(\varepsilon^2)\big) & \text{if } \alpha = 1,
\end{cases}
    \end{aligned}
\end{equation*}
which concludes the proof.
\end{proof}

%
%

\section{Stability of Euler and Heun Samplers}
\label{proof: euler-heun-stability}

\begin{proof}[Proof of Corollary \ref{cor: euler-heun-stability}]
As established in Appendices \ref{proof: euler-equivariance} and \ref{proof: heun-equivariance}, both the Euler and Heun samplers are permutation equivariant. We will proceed by showing that each sampler is state stable (Definition \ref{def: state-stable}) with a specific constant $\alpha$, and graph stable (Definition \ref{def: graph-stable}) with a bounding function $\beta$. Then, applying Theorem \ref{thm: discrete-flow-stability}, we will arrive at the respective bounds.\vspace{2pt}

\noindent\textbf{Euler sampler.} By the triangle inequality and Lipschitz continuity of the vector field with constant $M$, we have
\begin{equation*}
    \begin{aligned}
        \|T_k^{\boldsymbol{\theta}}(\mathbf{x}; \mathbf{S}) &- T_k^{\boldsymbol{\theta}}(\mathbf{y};\mathbf{S})\| \\
        &= \|\mathbf{x} - \mathbf{y} + h\big(u_k^{\boldsymbol{\theta}} (\mathbf{x};\mathbf{S}) - u_k^{\boldsymbol{\theta}} (\mathbf{y};\mathbf{S})\big)\|\\
        &\leq \|\mathbf{x}-\mathbf{y}\| + h\|u_k^{\boldsymbol{\theta}}(\mathbf{x};\mathbf{S})- u_k^{\boldsymbol{\theta}}(\mathbf{y};\mathbf{S})\| \\
        &\leq \|\mathbf{x}-\mathbf{y}\| + hM\|\mathbf{x}-\mathbf{y}\| \\
        & = (1+hM)\|\mathbf{x}-\mathbf{y}\|.
    \end{aligned}
\end{equation*}
Thus, the Euler sampler is state stable with $\alpha = 1+hM$.

For graph stability, we evaluate the error between the perturbed and nominal update maps using the stability of the continuous vector field (\ref{eq: VF-stability}), which yields
\begin{equation*}
    \begin{aligned}
        \|\mathbf{P}_0^\top T_k^{\boldsymbol{\theta}}(\mathbf{x}; \tilde{\mathbf{S}}) &- T_k^{\boldsymbol{\theta}}(\mathbf{P}_0^\top \mathbf{x};\mathbf{S})\| \\
        &= h\|\mathbf{P}_0^\top u_k^{\boldsymbol{\theta}} (\mathbf{x};\tilde{\mathbf{S}})- u_k^{\boldsymbol{\theta}}(\mathbf{P}_0^\top \mathbf{x};\mathbf{S})\|\\
        & \leq h \|g(\mathbf{x}, t_k)\| \big(\Gamma\varepsilon + \mathcal{O}(\varepsilon^2)\big).
    \end{aligned}
\end{equation*} 
We identify and define the bounding function $\beta(\mathbf{x}, t_k) \coloneqq h \|g(\mathbf{x}, t_k)\|$. To verify that $\beta$ is permutation \emph{invariant}, we use the permutation equivariance of $F$ (\ref{eq: temporal-conditioning-equivariance}) and the orthogonality of $\mathbf{P}^\top$, namely
\begin{equation*}
    \begin{aligned}
    \beta(\mathbf{P}^\top \mathbf{x}, t_k) 
    &= h \|g(\mathbf{P}^\top \mathbf{x}, t_k)\| = h \|\mathbf{P}^\top g(\mathbf{x}, t_k)\| \\
    & = h \|g(\mathbf{x}, t_k)\| = \beta(\mathbf{x}, t_k).
\end{aligned}
\end{equation*}
Thus, the Euler sampler is graph stable. Substituting $\alpha$ and $\beta_{\max} = h C_g(\mathbf{x}_0)$ into Theorem \ref{thm: discrete-flow-stability} concludes the proof.\vspace{2pt}

\noindent\textbf{Heun sampler.} Applying the exact same logic, we evaluate the state stability of the Heun update map. Using the triangle inequality and applying the Lipschitz constant $M$ sequentially
\begin{equation*}
    \begin{aligned}
        & \|T_k^{\boldsymbol{\theta}}(\mathbf{x}; \mathbf{S}) - T_k^{\boldsymbol{\theta}}(\mathbf{y};\mathbf{S})\| \\
        &\leq \|\mathbf{x}-\mathbf{y}\| + \frac{h}{2}\|u_k^{\boldsymbol{\theta}}(\mathbf{x};\mathbf{S}) - u_k^{\boldsymbol{\theta}}(\mathbf{y};\mathbf{S})\| \\
        &\quad + \frac{h}{2}\left\|u_{k+1}^{\boldsymbol{\theta}}\big(\mathbf{x} + hu_k^{\boldsymbol{\theta}}(\mathbf{x};\mathbf{S}); \mathbf{S}\big) - u_{k+1}^{\boldsymbol{\theta}}\big(\mathbf{y} + hu_k^{\boldsymbol{\theta}}(\mathbf{y};\mathbf{S}); \mathbf{S}\big)\right\|\\
        &\leq \|\mathbf{x}-\mathbf{y}\| + \frac{hM}{2}\|\mathbf{x}-\mathbf{y}\| \\
        &\quad + \frac{hM}{2}\Big(\|\mathbf{x}-\mathbf{y}\| + h\|u_k^{\boldsymbol{\theta}}(\mathbf{x};\mathbf{S}) - u_k^{\boldsymbol{\theta}}(\mathbf{y};\mathbf{S})\|\Big)\\
    &\leq \|\mathbf{x}-\mathbf{y}\| + \frac{hM}{2}\|\mathbf{x}-\mathbf{y}\| + \frac{hM}{2}\big(\|\mathbf{x}-\mathbf{y}\| + hM\|\mathbf{x}-\mathbf{y}\|\big) \\
        &= \left(1 + hM + \frac{h^2M^2}{2}\right) \|\mathbf{x}-\mathbf{y}\|.
    \end{aligned}
\end{equation*}
Thus, the Heun sampler is state stable with $\alpha = 1 + hM + \frac{h^2M^2}{2}$.

For graph stability, let $\tilde{\mathbf{z}} \coloneqq \mathbf{x} + hu_k^{\boldsymbol{\theta}}(\mathbf{x};\tilde{\mathbf{S}})$ and $\mathbf{z} \coloneqq \mathbf{P}_0^\top \mathbf{x} + hu_k^{\boldsymbol{\theta}}(\mathbf{P}_0^\top \mathbf{x};\mathbf{S})$. The error in the update map is bounded by
\begin{equation*}
    \begin{aligned}
        \|\mathbf{P}_0^\top T_k^{\boldsymbol{\theta}}(\mathbf{x}; \tilde{\mathbf{S}}) &- T_k^{\boldsymbol{\theta}}(\mathbf{P}_0^\top \mathbf{x};\mathbf{S})\| \\
        &\leq \frac{h}{2} \|\mathbf{P}_0^\top u_k^{\boldsymbol{\theta}} (\mathbf{x};\tilde{\mathbf{S}})- u_k^{\boldsymbol{\theta}}(\mathbf{P}_0^\top \mathbf{x};\mathbf{S})\| \\
        &\quad + \frac{h}{2} \|\mathbf{P}_0^\top u_{k+1}^{\boldsymbol{\theta}} (\tilde{\mathbf{z}};\tilde{\mathbf{S}})- u_{k+1}^{\boldsymbol{\theta}}(\mathbf{z};\mathbf{S})\|.
    \end{aligned}
\end{equation*}
The first term is bounded directly by $\frac{h}{2} \|g(\mathbf{x}, t_k)\| (\Gamma\varepsilon + \mathcal{O}(\varepsilon^2))$. For the second term, we add and subtract $\mathbf{P}_0^\top u_{k+1}^{\boldsymbol{\theta}}(\mathbf{P}_0 \mathbf{z}; \tilde{\mathbf{S}})$ and use the Lipschitz continuity of the vector field and its stability evaluated at $\mathbf{z}$ to obtain
\begin{equation*}
    \begin{aligned}
        \|\mathbf{P}_0^\top & u_{k+1}^{\boldsymbol{\theta}} (\tilde{\mathbf{z}};\tilde{\mathbf{S}}) - u_{k+1}^{\boldsymbol{\theta}}(\mathbf{z};\mathbf{S})\| \\
        &\leq \|\mathbf{P}_0^\top u_{k+1}^{\boldsymbol{\theta}} (\tilde{\mathbf{z}};\tilde{\mathbf{S}}) - \mathbf{P}_0^\top u_{k+1}^{\boldsymbol{\theta}} (\mathbf{P}_0 \mathbf{z};\tilde{\mathbf{S}})\| \\
        &\quad + \|\mathbf{P}_0^\top u_{k+1}^{\boldsymbol{\theta}} (\mathbf{P}_0 \mathbf{z};\tilde{\mathbf{S}}) - u_{k+1}^{\boldsymbol{\theta}}(\mathbf{z};\mathbf{S})\| \\
        &\leq M \|\tilde{\mathbf{z}} - \mathbf{P}_0 \mathbf{z}\| + \|g(\mathbf{z}, t_{k+1})\| \big(\Gamma\varepsilon + \mathcal{O}(\varepsilon^2)\big).
    \end{aligned}
\end{equation*}
Since $\|\tilde{\mathbf{z}} - \mathbf{P}_0 \mathbf{z}\| = h \|\mathbf{P}_0^\top u_k^{\boldsymbol{\theta}} (\mathbf{x};\tilde{\mathbf{S}})- u_k^{\boldsymbol{\theta}}(\mathbf{P}_0^\top \mathbf{x};\mathbf{S})\| \leq h \|g(\mathbf{x}, t_k)\| (\Gamma\varepsilon + \mathcal{O}(\varepsilon^2))$, we can combine everything to bound the total graph error by
\begin{equation*}
    \begin{aligned}
        \frac{h}{2} \Big( \|g(\mathbf{x}, t_k)\| + hM \|g(\mathbf{x}, t_k)\| + \|g(\mathbf{z}, t_{k+1})\| \Big) \big(\Gamma\varepsilon + \mathcal{O}(\varepsilon^2)\big).
    \end{aligned}
\end{equation*} 
Hence, the Heun sampler is graph stable with bounding function
\begin{equation}
    \beta(\mathbf{x}, t_k) \coloneqq \frac{h}{2} \Big( (1 + hM)\|g(\mathbf{x}, t_k)\| + \|g(\mathbf{z}, t_{k+1})\| \Big).
\end{equation}
By the definition of $C_g(\mathbf{x}_0)$, evaluated along the unperturbed trajectories, we have $\|g(\mathbf{x}_k, t_k)\| \leq C_g(\mathbf{x}_0)$ and $\|g(\mathbf{z}, t_{k+1})\| \leq C_g(\mathbf{x}_0)$. Therefore, evaluating the bounding function along this trajectory yields
\begin{equation*}
    \begin{aligned}
        \beta_{\max} &\leq \frac{h}{2} \Big( (1 + hM)C_g(\mathbf{x}_0) + C_g(\mathbf{x}_0) \Big) \\
        &= h \left(1 + \frac{hM}{2}\right) C_g(\mathbf{x}_0).
    \end{aligned}
\end{equation*}

Making the required substitutions in the stability bound of Theorem \ref{thm: discrete-flow-stability}, we find
\begin{equation*}
    \begin{aligned}
        & \min_{\mathbf{P} \in \mathcal{P}} \|\mathbf{P}^\top \Psi_k^{\boldsymbol{\theta}}(\mathbf{x}_0; \tilde{\mathbf{S}}) - \Psi_k^{\boldsymbol{\theta}}(\mathbf{P}^\top \mathbf{x}_0; \mathbf{S})\|\\ 
        &\leq \frac{\alpha^k - 1}{\alpha - 1} \beta_{\max} \big(\Gamma\varepsilon + \mathcal{O}(\varepsilon^2)\big)\\
        &\leq \frac{\left(1+ hM + \frac{h^2M^2}{2}\right)^k -1}{M} C_g(\mathbf{x}_0) \big(\Gamma\varepsilon  + \mathcal{O}(\varepsilon^2)\big).
    \end{aligned}
\end{equation*}
\end{proof}

%
%

\section{Wasserstein Stability of the Discrete Flow}
\label{proof: discrete-wasserstein-stability}

\begin{proof}[Proof of Corollary \ref{cor: discrete-wasserstein-stability}]
Let $\mathbf{P}_0 \in \mathcal{P}$ be the specific permutation matrix from Section \ref{sec: stability-of-gnn} that minimizes the relative perturbation error. Evaluating the discrete flow on the exact same initial random vector $\mathbf{x}_0$ under both the nominal graph $\mathbf{S}$ and the perturbed graph $\tilde{\mathbf{S}}$ forms a valid coupling. Thus, we can upper bound the Wasserstein-2 distance using the bound derived in Theorem \ref{thm: discrete-flow-stability}, to obtain
\begin{equation*}
    \begin{aligned}
        \min_{\mathbf{P} \in \mathcal{P}}W_2( \tilde{\pi}_k^{\mathbf{P}}, \pi_k^{\mathbf{P}} ) ^2 
        &\leq \mathbb{E}_{\mathbf{x}_0}\left[ \|\mathbf{P}^\top_0 \Psi_k^{\boldsymbol{\theta}}(\mathbf{x}_0;\tilde{\mathbf{S}}) - \Psi_k^{\boldsymbol{\theta}}(\mathbf{P}^\top_0 \mathbf{x}_0;\mathbf{S})\|^2 \right] \\
        &\leq \mathbb{E}_{\mathbf{x}_0}\left[ \Big(\Upsilon_k(\alpha) \beta_{\max}(\mathbf{x}_0) \big(\Gamma \varepsilon + \mathcal{O}(\varepsilon^2)\big)\Big)^2 \right].
    \end{aligned}
\end{equation*}

Because $\Upsilon_k(\alpha)$ and $\big(\Gamma \varepsilon + \mathcal{O}(\varepsilon^2)\big)$ are deterministic and independent of $\mathbf{x}_0$, we can factor them out of the expectation
\begin{equation*}
    \begin{aligned}
        \mathbb{E}_{\mathbf{x}_0} &\left[ \Big(\Upsilon_k(\alpha) \beta_{\max}(\mathbf{x}_0) \big(\Gamma \varepsilon + \mathcal{O}(\varepsilon^2)\big)\Big)^2 \right] \\
        &= \Upsilon_k(\alpha)^2 \big(\Gamma \varepsilon + \mathcal{O}(\varepsilon^2)\big)^2 \mathbb{E}_{\mathbf{x}_0}\left[ \beta_{\max}(\mathbf{x}_0)^2 \right] \\
        &= \Upsilon_k(\alpha)^2 \big(\Gamma \varepsilon + \mathcal{O}(\varepsilon^2)\big)^2 (\beta'_{\max})^2.
    \end{aligned}
\end{equation*}
Taking the square root of both sides directly yields \eqref{eq:bound_cor_4}.
\end{proof}

%
%

\section{Lipschitz constant of the vector field}
\label{proof: lipschitz-bound}

\begin{proof}[Proof of Proposition \ref{prop: lipschitz-bound}]
For a normalized Lipschitz pointwise nonlinearity $\sigma(\cdot)$, i.e., $|\sigma(b) - \sigma(a)| \leq |b-a|$, we have
\begin{equation*}
        \|\mathbf{X}_\ell - \mathbf{Y}_\ell\|^2_F 
        \leq \left \|\sum_{p=0}^{P-1} \mathbf{S}^p (\mathbf{X}_{\ell-1} - \mathbf{Y}_{\ell-1}) \boldsymbol{\Theta}_{\ell p}\right\|_F^2.
\end{equation*}
Let $\mathbf{Z}_{\ell-1} \coloneqq \mathbf{X}_{\ell-1} - \mathbf{Y}_{\ell-1}$ and define its GFT as $\bar{\mathbf{Z}}_{\ell-1} \coloneqq \mathbf{V}^\top \mathbf{Z}_{\ell-1}$. Substituting $\mathbf{S} = \mathbf{V}\mathbf{\Lambda}\mathbf{V}^\top$ yields
\begin{equation*}
        \|\mathbf{X}_\ell - \mathbf{Y}_\ell\|^2_F 
        \leq \left \|\sum_{p=0}^{P-1} \mathbf{\Lambda}^p\bar{\mathbf{Z}}_{\ell -1} \boldsymbol{\Theta}_{\ell p}\right\|_F^2.
\end{equation*}
Let $\bar{\mathbf{Y}} \coloneqq \sum_{p=0}^{P-1} \mathbf{\Lambda}^p\bar{\mathbf{Z}}_{\ell -1} \boldsymbol{\Theta}_{\ell p}$ and denote its $i$-th row as $\bar{\mathbf{y}}_i^\top$. Because $\mathbf{\Lambda}=\textrm{diag}(\lambda_1,\ldots,\lambda_N)$, the operations decouple across the rows in the spectral domain
\begin{align*}
    \bar{\mathbf{y}}_i^\top = \sum_{p=0}^{P-1} \lambda_i^p \bar{\mathbf{z}}_i^\top \boldsymbol{\Theta}_{\ell p} = \bar{\mathbf{z}}_i^\top \mathbf{H}_\ell(\lambda_i),
\end{align*}
where $\mathbf{H}_\ell(\lambda_i) \coloneqq \sum_{p=0}^{P-1} \lambda_i^p \boldsymbol{\Theta}_{\ell p}$. Then
\begin{equation*}
    \begin{aligned}
        \|\mathbf{X}_\ell &- \mathbf{Y}_\ell\|^2_F \leq  \left \|\bar{\mathbf{Y}}\right\|_F^2 = \sum_{i=1}^N \|\bar{\mathbf{y}}_i\|^2 = \sum_{i=1}^N \|\bar{\mathbf{z}}_i^\top \mathbf{H}_\ell(\lambda_i)\|^2 \\
        & \leq \sum_{i=1}^N \|\bar{\mathbf{z}}_i\|^2 \|\mathbf{H}_\ell(\lambda_i)\|^2_2 \leq M_\ell^2 \sum_{i=1}^N \|\bar{\mathbf{z}}_i\|^2\\
        &= M_\ell^2 \|\bar{\mathbf{Z}}_{\ell-1}\|^2_F = M_\ell^2 \|\mathbf{X}_{\ell-1}- \mathbf{Y}_{\ell -1}\|_F^2,
    \end{aligned}
\end{equation*}
where $M_\ell$ is the layer-wise Lipschitz constant defined as
\begin{align*}
    M_\ell \coloneqq \max_{i \in \{1, \dots, N\}} \left\| \sum_{p=0}^{P-1} \boldsymbol{\Theta}_{\ell p} \lambda_i^p \right\|_2.
\end{align*}
Applying this relation recursively across all $L$ layers bounds the global spatial Lipschitz constant $M \leq \prod_{\ell=1}^L M_\ell$, which concludes the proof. 
\end{proof}



\bibliographystyle{IEEEtran}
\bibliography{references}



\end{document}